\documentclass[prd,showpacs,nofootinbib,preprintnumbers]{revtex4}

\usepackage{latexsym}
\usepackage{amssymb}
\usepackage{amsmath}
\usepackage{slashbox}
\usepackage{subfigure}
\usepackage{hyperref}
\usepackage[russian,english]{babel}
\usepackage[dvips]{epsfig}
\numberwithin{equation}{section}

\newcommand{\ull}{p_T}
\newcommand{\uln}{k_T}
\newcommand{\ga}{\gamma}\newcommand{\om}{\omega}
\newcommand{\del}{{\delta}}
\newcommand{\lbr}{\left(}
\newcommand{\rbr}{\right)}
\newcommand{\cp}{{\cal{P}}}
\newcommand{\sg}{{\sqrt{-g}}}

\newcommand{\ka}{\varkappa_D}
\newcommand{\vk}{\varkappa_D}
\newcommand{\pa}{\partial}
\newcommand{\fr}{\frac}

\newcommand{\lb}{\label}
\newcommand{\be}{\begin{equation}}
\newcommand{\ee}{\end{equation}}
\newcommand{\ba}{\begin{align}}
\newcommand{\ea}{\end{align}}
\newcommand{\bea}{\begin{eqnarray}}
\newcommand{\eea}{\end{eqnarray}}
\newcommand{\bw}{\begin{widetext}}
\newcommand{\ew}{\end{widetext}}

\newcommand{\ep}{{\varepsilon}}
\newcommand{\ffi}{{\varphi}}

\newcommand{\e}{{\rm e}}
\newcommand{\al}{{\alpha}}

\newcommand{\nn}{\nonumber}
\newcommand{\zt}{\dot{z}}
\newcommand{\Hh}{h}

\newcommand{\cd}{,\! \,}
\newcommand{\un}{\, ^1  }
\newcommand{\nul}{\, ^0 }
\newcommand{\de}{\, ^2 }
\newcommand{\pp}{\, ... \,}
\newcommand{\y}{{\mathbf{y}}}

\newcommand{\od}{\omega}

\newcommand{\coa}{\xi}
\newcommand{\comment}[1]{}

\newcommand{\tD}{\Gamma}
\newcommand{\bl}{\vphantom{u'}}
\newcommand{\ths}{}

\newcommand{\ds}{\displaystyle}
\newcommand{\fip}{ \varepsilon_{\rm I}^{MN}}

\newcommand{\LW}{{Li\'{e}nard-Wiechert }}
\newcommand{\dtilde}[1]{\tilde{\tilde{#1}}}

\begin{document}

\hfill CCTP-2012-21

\hfill CERN-PH-TH/2012-275

\title{Gravitational  bremsstrahlung in ultra-planckian collisions}
\author{Dmitry Gal'tsov$^a$, Pavel Spirin$^a$ and Theodore N. Tomaras$^{b,c}$
\thanks{\tt E-mail: galtsov@physics.msu.ru,
salotop@list.ru, tomaras@physics.uoc.gr}} \affiliation{
 \mbox{$^a$Department of Theoretical
Physics, Moscow State University, 119899, Moscow, Russia;}\\
\mbox{$^b$CERN, Theory Division;} \\
 \mbox{$^c$Department of Physics and Crete Center for Theoretical Physics,}\\
 \mbox{\hspace{0.5cm} University of Crete, 71003, Heraklion, Greece.}}

\date{\today}

\begin{abstract}
A classical computation of gravitational bremsstrahlung in
ultra-planckian collisions of massive point particles is presented
in an arbitrary number $d$ of toroidal or non-compact extra
dimensions. Our method generalizes the post-linear formalism of
General Relativity to the multidimensional case. The total emitted
energy, as well as its angular and frequency distribution are
discussed in detail. In terms of the gravitational radius $r_S$ of
the collision energy, the impact parameter $b$ and the Lorentz
factor in the CM frame, the leading order radiation efficiency in
the Lab frame is shown to be $\epsilon\sim (r_S/b)^{3(d+1)}
\gamma_{\rm cm}$ for $d=0, 1$ and $\epsilon\sim (r_S/b)^{3(d+1)}
\gamma_{\rm cm}^{\;2d-3}$ for $d\geqslant 2$, up to a known
$d$-dependent coefficient and a $\ln\gamma_{\rm cm}$ factor for
$d=2$, while the characteristic frequency of the radiation is
$\omega\sim \gamma/b$. The contribution of the low frequency part
of the radiation (soft gravitons) to the total radiated energy is
shown to be negligible for all values of $d$. The domain of
validity of the classical result is discussed. Finally, it is
shown that within the region of validity of our approach the
efficiency can obtain unnatural values greater than one, which is
interpreted to mean that the peripheral ultra-planckian collisions
should be strongly radiation damped.
\end{abstract}\maketitle

\tableofcontents

\section{Introduction}\label{intr}
The scenario of TeV-scale gravity with large extra dimensions
\cite{ADD1,GRW} is currently being tested experimentally at the LHC
\cite{LHC}. In this class of higher-dimensional gravity theories the
Planck mass can be of the order of a TeV, provided that the number
of extra dimensions is greater than two. If so, the LHC will probe
ultra-planckian physics where gravity not only is the dominant force
\cite{'tHooft}, but in addition it is believed to be adequately
described by the classical Einstein equations \cite{GiRaWeTrans}.
This, presumably, allows one to make reliable theoretical
predictions of gravitational effects without entering into the
complications related to quantum gravity. One such prediction,
namely the possibility of black hole formation at the LHC was
extensively discussed in the literature during the past ten years
\cite{BH,reviews,LHC,BHrev,Eardley}.

The main focus of this paper is the study of gravitational bremsstrahlung in ultra-planckian particle collisions.
In ultra-planckian collisions with impact parameters smaller than the
Schwarzschild radius of the center of mass energy of the colliding particles,
the creation of a black hole is a dominant process and is
accompanied by radiation losses at the level of 10-40\%
depending on the number of extra dimensions (see
\cite{Coelho:2012sy} generalizing earlier calculations by D'Eath
and Payne \cite{D'Eath:1976ri, D'Eath:1992hb}). On the other hand, for impact
parameters larger than the gravitational radius, the black
hole formation process is exponentially suppressed and the gravitational
bremsstrahlung becomes the dominant inelastic classical effect, which furthermore
can be very strong due to the huge number of light Kaluza-Klein states in the graviton
spectrum. It is worth noting, that gravitational radiation may not be
maximal in head-on collisions. For instance, radiation from point particles
falling into the black hole is {\em minimal} for radial infall
and grows initially with the impact parameter, a fact which is well known
both in four-dimensional general relativity and in models with
extra dimensions. The way this is understood is the following: ultra-relativistic particles hitting with
some critical impact parameter can be captured into quasi-circular
orbits before they plunge into the hole. During these revolutions they radiate with almost constant rate
and lose energy more efficiently than during a radial infall.

A large number of papers from various groups is devoted to the
study of gravitational bremsstrahlung in a variety of physical
set-ups and using different approaches. A rather incomplete list
includes the following: A classical calculation relevant to zero
and small impact parameter using the colliding wave picture was
suggested long ago by D'Eath \cite{D'Eath:1976ri, D'Eath:1992hb}.
More recently this approach was extended to the extradimensional
theories \cite{Eardley, Coelho:2012sy}. The approach of Smarr et
al. in \cite{Smarr:1976qy} was based on the low frequency
approximation for the amplitude to estimate the full energy loss.
Matzner and Nutku proposed the use of the method of virtual
gravitons \cite{Matzner:1974rd} in analogy with the
Weizs\"acker-Williams approach in quantum electrodynamics.

Another approach consists in considering radiation in the
linearized theory from particles falling towards a black hole
\cite{Cardlemos, Cardoso:2005jq}. Gravitational bremsstrahlung was
also studied in the Born approximation of quantum gravity (in four
dimensions) \cite{Barker}, while the relationship between the Born
approximation and the post-linear formalism was discussed in
\cite{Galtsov:1980ap}. More recently, a novel approach, making use
of perturbative quantum gravity to derive effective equations of
motion and to describe gravitational radiation using those, was
proposed under the name of ``effective theory'' approach in
\cite{effective}. Estimates of radiation losses in the eikonal
approximation of (multidimensional) quantum gravity were given in
\cite{GiRaWeTrans}, while Amati, Ciafaloni and Veneziano in a
series of papers \cite{ACV1993} have considered the problem of
radiation in ultra-high energy collisions of massless particles in
the context of quantum string theory.   Estimates of the
gravitational bremsstrahlung in high energy collisions in the ADD
model were given in \cite{Koch:2008zza}.   Finally, a lot of
numerical work has also been devoted to this subject
\cite{Anninos:1993zj,Choptuik:2009ww,Sperhake:2008ga,Yoshino,Yoshino3,MaOn}.

In four-dimensional theories, in particular, gravitational bremsstrahlung in small-angle ultrarelativistic scattering of
point particles was calculated long ago, most notably by Kovacs and Thorne \cite{KT}, in the framework
of the so-called ``fast motion approximation'' scheme, first proposed by Bertotti, Havas
and Goldberg \cite{FMA}. A similar independent calculation, using momentum
space perturbation theory up to second order in the gravitational
constant, was performed in Ref. \cite{Galtsov:1980ap} leading to
essentially the same results. The latter approach amounts to performing a perturbation expansion of the
metric around the Minkowski background up to second order in the gravitational constant and
is applicable for arbitrary mass-ratios of the colliding
particles. In the case of one mass $M$ much larger than the other $m$, the same result had been obtained
earlier by Peters, also using a linear approximation but around the Schwarzschild background of the heavy mass
\cite{Peters:1970mx}.

The purpose of this paper is to study gravitational radiation losses in the case of ultra-relativistic collisions
with large impact parameters, for which the scattering angle is small and
calculations can be performed reliably within the post-linear approximation
scheme of General Relativity. This paper continues a series of investigations
\cite{GKST-2,GKST-3,GKST-PLB}  devoted to bremsstrahlung in flat
space-time arising under non-gravitational scattering of charged
particles \cite{GKST-2} and scalar radiation in gravity-mediated particle collisions \cite{GKST-3}.
Here we focus on the problem of gravitational radiation in collisions of {\it massive} point
particles interacting gravitationally. The computation is purely classical and iterative and, as such, it can
only be reliable within a certain domain of validity in the space of parameters. Somewhat unexpectedly, it
turns out that, for parameter values within the region of validity of the classical approximation
the {\it radiation efficiency}, i.e. the fraction of the initial energy which is radiated away, is
substantially enhanced compared to the four-dimensional case,
contrary to earlier qualitative estimates \cite{Mironov:2006wi}.
Moreover, for a number of extra dimensions greater than two,
this quantity grows with a dimension-dependent power of the Lorentz-factor $\gamma$ of the collision.

Specifically, for the leading order radiation efficiency in the laboratory frame we obtain (up to a dimension-dependent
numerical coefficient, which however can be quite large, and an inessential $\ln\gamma$ factor for
$d=2$ extra dimensions)
\footnote{Notice that these expressions differ from the generic formula $\epsilon\sim (r_S/b)^{3(d+1)}
\gamma_{\rm cm}^{2d+1}$ given erroneously for all dimensions in \cite{GKST-PLB}.
The difference is due to an error in \cite{GKST-PLB}, related to the extent of the phenomenon of
destructive interference discussed in the text.
Nevertheless, the qualitative conclusions of \cite{GKST-PLB} are still correct and will be discussed
further in Section 5 of the present paper.}
\begin{equation} \epsilon\equiv\frac{E}{{\mathcal E}_0} \sim \left(\frac{r_S}{b}\right)^{3(d+1)} \gamma_{\rm cm}\;, \quad d=0,1
\end{equation}
\begin{equation} \epsilon\sim \left(\frac{r_S}{b}\right)^{3(d+1)} \gamma_{\rm cm}^{\; 2d-3}\;, \quad d\geqslant 2 \;,
\end{equation}
where $b$ is the impact parameter of the collision,
$r_S$ is the Schwarzschild radius for the center-of-mass collision energy
\begin{equation}
r_S=\frac{1}{\sqrt{\pi}} \left[ \frac{8\Gamma
\left( \frac{d+3}{2}\right)}{d+2}\right]^{
\frac{1}{d+1}}\left(\frac{G_D
 \sqrt{s}}{c^4}\right)^{\frac1{d+1}}\!\!,
\end{equation}
and ${\mathcal E}_0=m(\gamma+1) \simeq m\gamma$ is the initial energy in the Lab frame. For collisions
of two equal mass particles the Lorentz factors in the Lab and in the center-of-mass frames
are related by $\gamma = 2\gamma_{\rm cm}^2-1$.

In the special case of $d=0$ our results for the total emitted energy as well as for its angular and frequency distribution
are in perfect agreement with the results of Kovacs and Thorne \cite{KT}.

The present
paper is organized in five sections of which this Introduction is the first. In Section 2 the model, the basic formulae
and the iterative
procedure, that will be employed, are described. Also, a complete orthonormal set of graviton polarization tensors in
arbitrary dimensions, appropriate for the problem at hand are explicitly constructed. In Section 3
the expressions of the local and non-local contributions to the gravitational radiation source are obtained. The
phenomenon of destructive interference of the leading local and non-local terms is shown. The total radiation amplitude is
also obtained. Section 4 contains the computation of the leading ultra-relativistic order emitted energy in the
classical theory in arbitrary dimensions.
It also contains a detailed discussion of the frequency and angular distributions of the emitted
radiation. The zero frequency limit of the radiation is analyzed in detail in all dimensions.
In the same Section the domain of validity of the classical computation is discussed. The final result for
the radiation efficiency is given and the unnatural possibility for it, to take values greater than one and
even diverge in the massless limit for $d\geqslant 2$ extra dimensions, is
critically analyzed. A summary together with a few final comments are given in the final Conclusion section.

\section{General setting}

We will study gravitational bremsstrahlung both in uncompactified
Minkowski space-time $M_{1,D-1}=M_{1,3}\times \mathbb{R}^d$ and in
the ADD model $M_{1,3}\times T^d$ with the $d$ extra dimensions
compactified on a torus. In both cases the collision will be
confined on a brane $M_{1,3}$, and the $D-$dimensional cartesian
coordinates are split as $x^M=(x^{\mu};y^i)$, $x^{\mu} \in
M_{1,3}$, $y^i \in \mathbb{R}^d$ or $T^d$, respectively. The brane
subspace $M_{1,3}$ is selected by the initial conditions on the
particle velocities and is fixed throughout the collision process,
since the gravitational interaction cannot expel the particles
from the brane.

\subsection{The model}

Consider two point masses $m$ and $m'$ moving along the world-lines
$x^M=z^M(\tau)$ and $x^M=z'^M(\tau')$ ($M=0,1,\ldots,D-1$) and interacting with the $D-$dimensional
gravitational field. The corresponding action is
\begin{equation}
\label{actiongk} S=-\sum \frac{1}{2} \int \lbr e\; g_{MN}\dot{z}^M
\dot{z}^N+\frac{m^2}{e}\rbr\;d\tau    -\frac{1}{\ka^2}\int R_D
\,\sqrt{-g}\; d^D x \,,
 \end{equation}
where $\ka^2\equiv 16\pi G_D$, $e(\tau)$ is the ein-bein of the
trajectory and the summation is over the two particles. The metric
signature is chosen to be $(+,-,...,-)$ and our convention for the
Riemann tensor is $R^B{}_{NRS}\equiv \Gamma^B_{NS , R} -
\Gamma^B_{NR , S} + \Gamma^A_{NS} \Gamma^B_{AR} - \Gamma^A_{NR}
\Gamma^B_{AS}$, with $\Gamma^A_{NR}=(1/2)g^{AB}(g_{BR, N}+g_{N B ,
R}-g_{NR, B})$. Finally, the Ricci tensor and curvature scalar are
$R_{MN}\equiv \delta^B_A\, R^A{}_{MBN}$ and $R\equiv g^{MN}
\,R_{MN}$, respectively.

The action is invariant under the general coordinate reparametrization of spacetime, as well as under the
independent reparametrizations of the particle trajectories. For the trajectory of particle $m$ the
transformation is $\tau\to \tilde\tau=\tilde\tau(\tau)$,
under which all fields are scalars except for the einbein $e(\tau)$, which transforms according to
$e(\tau)\to \tilde e(\tilde\tau)=e(\tau) d\tilde\tau/d\tau$.

Varying $S$ with respect to $e(\tau)$ and $z^M(\tau)$ one obtains the equations
\begin{equation}
\label{consp}
e^2  g_{MN} \dot{z}^M \dot{z}^N=m^2
\end{equation}
and
\begin{equation}\label{eomp}
\fr{d}{d\tau}\lbr e \dot{z}^N g_{MN} \rbr=\fr{e}2 \;
g_{NP,M}\dot{z}^N \dot{z}^P ,
\end{equation}
respectively. Two analogous equations are obtained by varying with respect to $e'$ and ${z'}^M$,
while variation of $g_{MN}$ leads to the Einstein equations
\begin{equation}
\label{Eeq} G^{MN}=\frac{1}{2}\,\ka^2 T^{MN},\quad T^{MN}=\sum e
\int \fr{\zt^M \zt^N \del^D(x-z(\tau))}{\sg}d\tau \,.
\end{equation}

Strictly speaking  the notion of point-like particles is not compatible with full non-linear gravity.
Nevertheless, it still makes sense in the context of the weak-field perturbation expansion around
flat spacetime, which will be adopted here. In this approach one writes the metric as
\begin{equation}
\label{meka}
g_{MN}=\eta_{MN}+\ka h_{MN}\,,
\end{equation}
and expands all quantities in powers of $h_{MN}$, using $\eta_{MN}$ to raise and lower the indices.
The flat background will be either $D$-dimensional Minkowski space $M_{1, D-1}$, or the product
$M_{1,3}\times T^d$ ($d=D-4$) of four-dimensional Minkowski and a $d-$torus.

It is convenient to define the quantity
\begin{equation}
\psi_{MN}\equiv h_{MN}-\fr12 h \, \eta_{MN}\,, \qquad h\equiv \eta^{MN}\,h_{MN}\,,
\end{equation}
and fix the general coordinate reparametrization
symmetry by choosing to work in the flat-space harmonic gauge
\begin{equation}
\label{hagef} \pa_N \psi^{MN}=0\,.
\end{equation}

The reparametrization freedom of the particle trajectories will be dealt with later.
The expansion of several relevant quantities in powers of $h_{MN}$ is given in Appendix \ref{app1}.
In particular, the Einstein tensor in the harmonic gauge becomes
\begin{equation}
\label{SMN}
G_{MN}=-\fr{\ka}2 \Box \psi_{MN}-\fr{\ka^2}2 S_{MN}+N_{MN},
\end{equation}
where $\Box=\pa_M\pa^M$ is the flat d'Alembert operator, $S_{MN}$ is the $\mathcal{O}(h^2)$ part
of $G_{MN}$ given by
\begin{align}
\label{natag_0} S_{MN} (\Hh) =& {\Hh}_M^{P \cd Q}(\Hh_{NQ \cd P} -
\Hh_{NP \cd Q}) +\Hh^{PQ}(\Hh_{MP \cd NQ}+ \Hh_{NP \cd MQ}-
\Hh_{PQ\cd MN}- \Hh_{MN
\cd PQ}) -  \nn\\
  -&\frac{1}{2} \Hh^{PQ}_{\quad \cd M} \Hh_{PQ \cd
N}-\frac{1}{2}\Hh_{MN}\Box \Hh +
\frac{1}{2}\eta_{MN}\left(2\Hh^{PQ}\Box \Hh_{PQ}-\Hh_{PQ \cd L}
\Hh^{PL \cd Q}+\frac{3}{2} \Hh_{PQ \cd L} \Hh^{PQ\cd L}\right),
\end{align}
while $N_{MN}$ stands for all cubic and higher in $h_{MN}$ terms in the Einstein tensor.

A few remarks are in order. Usually  in the ADD scenario one
assumes that matter is localized on the brane, i.e. restricted to
the subspace $M_{1,3}$. This assumption is non-contradictory within
the linearized gravity, since the matter energy momentum in this
case is purely flat-space tensor, whose conservation does not
involve gravity. In the full non-linear gravity the $D$-dimensional
Bianchi identity
\begin{align}\label{cons}
 T_{M; N}^{N}=\frac{1}{\sqrt{-g}}[T_{M}^{N} \sqrt{-g}]_{,N} - \frac{1}{2} \, g_{NP , M}
 T^{NP}=0
\end{align}
implies $D$-dimensional  geodesic equations for the particles
(\ref{eomp}) (and similarly for brane fields),  which are generally
inconsistent with the assumption of matter localization on the
brane. For this reason we have to consider the world-lines in
(\ref{actiongk}) from the beginning as $D$-dimensional curves. In general, without any
extra forces ensuring confinement, the particle motion can be
localized on the brane only by suitable initial conditions. However, it will be shown
that the gravitational interaction treated perturbatively
will not expel the colliding particles from the brane at least to leading order in our iterative scheme.
This will be sufficient for the consistency of our computation of bremsstrahlung.

Another technical difference from the standard ADD approach, also related
to the fact that our treatment appeals to the $D$-dimensional
picture rather than to  the four-dimensional one, will be the
treatment of the graviton as a $D$-dimensional massless particle, instead
of as a set of four-dimensional massive fields of spins $2,\,1,\,0$ by suitable rearrangement of the
components of the $D$-dimensional graviton \cite{GRW}. In particular, it will be more convenient here
to choose directly the $D(D-3)/2$ polarization tensors of
the graviton propagating in the bulk, rather than to split them into
polarizations of massive particles in four dimensions.

\subsection{The iteration scheme}

Our approach is identical to the one used in simpler models in \cite{GKST-2} and \cite{GKST-3}.
The iterative solution of the above field equations and gauge fixing conditions amounts to
formally expanding and computing all fields step-by-step in the form
\begin{equation}
\Phi=\nul\Phi+\un\Phi+{^2}\Phi+\ldots\,,
\end{equation}
 where
$\Phi$ denotes any of the fields $z^M(\tau)$, ${z'}^M(\tau')$,
$e(\tau)$, $e'(\tau')$ and $h_{MN}(x)$ and with the left
superscript labeling the step of the iteration.

\vspace{0.3cm}

{\it The zeroth order solution} is trivial. It describes the two
particles moving with constant velocities ($u^M=(u^{\mu},
0, ..., 0)$ and $u'^M=(u'^{\mu}, 0, ..., 0)$, respectively in
the $M_{1,3}$ subspace, whose indices are labeled by lower case
greek letters $\mu, \nu=0, 1, 2, 3$. Correspondingly, their trajectories
$\nul z^M=(\nul z^\mu(\tau), 0..., 0)$, and similarly for $\nul z'^M$,
are the straight lines
\begin{equation}
\nul z^\mu(\tau)=z^\mu(0)+u^\mu\tau \,, \quad \nul
{z'}^\mu(\tau')={z'}^\mu(0)+{u'}^\mu\tau' ,
\end{equation}
in the absence of any gravitational field
\begin{equation} \nul
h_{MN}=0\,.
\end{equation}
 The Lagrange multipliers are chosen
equal to the corresponding particle masses
\begin{equation}\nul e =m\,, \;\; \nul e' =m' ,
\end{equation}
 so that the trajectories are parametrized by the corresponding
proper times and the four-velocities satisfy the normalization
conditions $\eta_{MN}u^M u^N\equiv u^2=u'^2=1$.

We choose to work in the Lorentz frame in which the target
particle $m'$ is initially at rest and with the projectile $m$
moving along the $z$-axis. Also, we introduce the space-like
vector $b^M=(b^{\mu}, 0, ...,0)$ lying on the brane, with:
\begin{equation}
b^{\,\mu}=z^\mu(0)-{z'}^\mu(0)\,.
\end{equation}
With no loss of generality one can further assume $(b u)=(b u')=0$ and choose the $x$-axis along $\mathbf{b}$.
Thus,
\begin{equation}
 \label{u}
 u^\mu=\gamma(1,0,0,v)\,,\qquad
{u'}^\mu=(1,0,0,0)\,, \qquad b^{\,\mu}=(0,b,0,0)\,,
 \end{equation}
 where $\gamma=1/\sqrt{1-v^2}$ and $b$ is the impact parameter.

 \vspace{0.3cm}

{\it The first order correction} is obtained next. The zeroth order
straight particle trajectories are sources of the first order
gravitational field $\un h_{MN} (\un h'_{MN})$ of each particle,
which in turn, causes the first order deviation of the trajectory
$\un {z'}^M (\un z^M)$ of the other one. In the process, the first
correction $\un e (\un e')$ of the einbein fields is also
obtained. Explicitly, from the zeroth order trajectories one
obtains the zeroth order energy-momentum tensor
\begin{equation}
\label{T0mn} \nul T^{MN}=\sum m\int u^M u^N \delta^D(x-\nul
z(\tau))\, d\tau\,,
\end{equation}
which in this order has only $\{M, N\}=\{\mu, \nu\}$ components, and from the first order Einstein equations,
given in the harmonic gauge by
\begin{equation}
\label{hanuleq}
\Box\, \un\psi^{MN}=- \ka \nul T^{MN},
\end{equation}
the first order correction $\un\psi^{MN}$ to the metric is
obtained. Although only the $\un\psi^{\mu\nu}$ components of
$\un\psi^{MN}$ are non-vanishing, the corresponding $\un h^{MN}$
will also have non-vanishing bulk components from the trace part
in Eq. (\ref{hanuleq}).

Using $\un h^{MN}$ and the zeroth order solution in equations (\ref{consp}) and (\ref{eomp}) one
obtains for $\un e$ and $\un z^M$ the equations
\footnote{Our gauge condition is $g_{MN} \dot z^M \dot z^N=1$. To this order it reduces to $^1e=0$.}
\begin{equation}
\label{e1eq} \un e=-\frac{ m}{2} \lbr \ka \un
h_{MN}u^Mu^N+2\,\eta_{MN}u^M \un
 \dot{z}^N \rbr
\end{equation}
and
\begin{equation}
\fr{d}{d\tau}\lbr \un e u_M +m \un \dot{z}_M\rbr=-\ka m
H_{PQM}\;u^P u^Q,\qquad H_{PQM}=\un h_{PM,Q}-\fr12 \un h_{PQ,M}\,,
\end{equation} which upon elimination of $\un e$ give for $\un z^M$
\begin{equation}
\label{z1eq}
\Pi^{MN} \fr{d}{d\tau} \un\dot{z}_N=-\ka\Pi^{MN}H_{PQN}\;u^P
u^Q \,,
\end{equation}
where
\begin{equation}
\Pi^{MN}=\eta^{MN}-u^M u^N
\end{equation}
is the projector onto the subspace orthogonal to $u^M$. Equations (\ref{hanuleq}), (\ref{e1eq}) and
(\ref{z1eq}), together with two similar equations
for $\un e'$ and $\un {z'}^M$, form a complete set of equations in this order. The gauge fixing condition
(\ref{hagef}) for $\un \psi^{MN}$ is a consequence of the conservation of $\nul T^{MN}$.

It should be kept in mind that in the equation of motion (\ref{z1eq}) of particle $m$
the gravitational field on the right hand side is the one due to $m'$.
Singularities due to the action on $m$ of its own gravitational field can be removed in lowest
order by classical renormalization of the affine parameter of its trajectory \cite{GS}; this will be enough
for our purposes. In what follows we will omit self-action terms and consider only mutual gravitational
interaction.

\vspace{0.3cm}

\begin{figure}
\begin{center}
\includegraphics[angle=0,width=15cm]{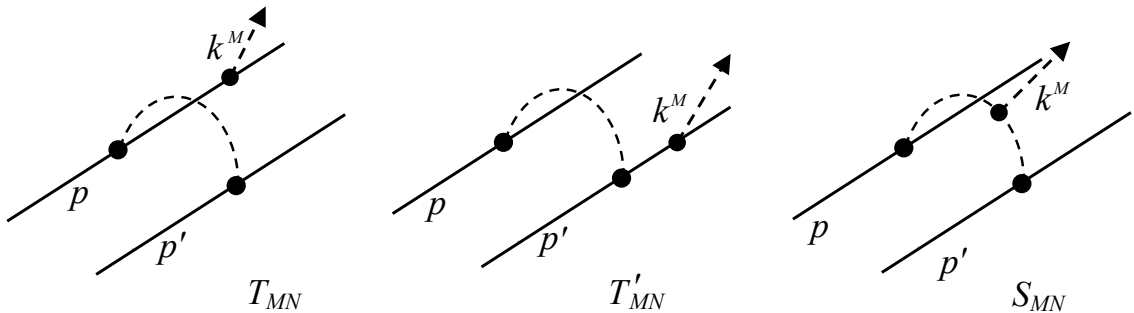}
\caption{Schematic representation of the sources of gravitational radiation described by (\ref{psi2eq}). The
interactions are meant to all orders in the quantum perturbation expansion sense. Solid (dotted) lines
are colliding particles (gravitons).}
\label{diagrams}
\end{center}
\end{figure}

{\it In the next order of iteration} one obtains the leading contribution to gravitational radiation. It is due
to the accelerated particles, as well as to the cubic gravitational self-interaction. Indeed,
the Einstein field equation reduces for the second order correction $^2\psi_{MN}$ of
the gravitational field to
\begin{equation}
\label{psi2eq} \Box\, \de\psi_{MN}=-\ka\;\tau_{MN}\,,
\end{equation}
with the source term having three contributions:
\begin{equation}\label{source2}
\tau_{MN}= \un T_{MN}+\un T'_{MN}+S_{MN}\lbr \un h \rbr .
\end{equation}
The first two terms of $\tau_{MN}$ (\ref{source2}) represent the
particles' contribution to radiation. As obtained to this order
from (\ref{Eeq}) they are \footnote{Symmetrization over two
indices is defined according to $A_{(MN)}\equiv
(A_{MN}+A_{NM})/2$.}
 \begin{align}
\label{T1MN} \un T_{MN}(x)=  m \int   \left[ 2\un \dot z_{(M}
u_{N)} +\ka \lbr 2u^P \un h_{P(M}u_{N)} -\fr{\un h}2 u_M u_N\rbr -
u_M u_N \un {z}^P\pa_P \right] \delta^D\!(x-\! \! \nul z(\tau))\,
d\tau\,. \end{align} Ditto for $m'$ with $u$  replaced by $u'$. To
repeat, in (\ref{T1MN}) $\un h$ is due to $m'$. Finally, the part $S_{MN}$ of $\tau_{MN}$ represents the
contribution of the gravitational field itself to gravitational
radiation. Given that $S_{MN}$ is quadratic in $h_{MN}$,
consistency of the iteration to this order requires to substitute
$h_{MN} \to \un h_{MN}$ inside $S$. Furthermore, just as the
self-interaction terms were ignored in the particles' equations of
motion, one has to keep in $S_{MN}(\un h)$ only the products of
the first order corrections $\un h$ {\em due to different
particles}. The various contributions to the gravitational
radiation to this order is shown schematically in Figure
\ref{diagrams}.

It is straightforward to verify on the basis of the equations of motion satisfied by the first order fields that
\begin{equation}
\label{divtau}
\pa_N \tau^{MN}=0\,,
\end{equation}
which guarantees the validity of the gauge fixing condition (\ref{hagef}) to this order.

So $S_{MN}$ considered as a quadratic form in $\un h_{MN}$
constitutes the non-local (in terms of the flat space picture)
source of gravitational radiation. This non-locality is due to the
non-linearity of the theory and, as will be shown, leads to
important differences in the radiation spectrum from linear
theories like electromagnetism. More detailed discussions of this
point in four dimensions can be found in \cite{Galtsov:1980ap}, as
well as in its generalization to arbitrary dimension but in a
simpler non-linear model in \cite{GKST-3}.

\subsubsection{The asymptotic behavior of the fields}

The same tensor $S_{MN}$ considered as a quadratic form  in $\de
h_{MN}$ plays the role of an effective energy-momentum tensor of
gravitational waves \cite{Weinberg}. This interpretation is based
on the conservation equation (\ref{divtau}) in combination with
the asymptotic properties of the solutions of the flat-space
d'Alembert equation and is valid in all space-time dimensions.

Recall, that in four dimensions $\un h_{\mu\nu}$, which according to (\ref{hanuleq}) is due of a static or a
uniformly moving mass, falls-off as
$1/r$, and the corresponding ``field strength'' (derivatives $\un h_{\mu\nu,\lambda}$) behave as $1/r^2$.
On the other hand, the fall-off of the radiative component of the retarded potential $\de h_{\mu\nu}$ is $1/r$ and
the same for its derivatives $\de
h_{\mu\nu,\lambda}$. Correspondingly, in $D$-dimensional Minkowski
space-time the behaviors of $\un h_{MN}$ and its derivatives
$\un h_{MN,L}$ are $1/r^{D-3}$ and $1/r^{D-2 }$, respectively,
where $r$ is the radial coordinate in the $D-1$-dimensional space.
The corresponding fall-off for the radiation field can be found
from the recurrence relations for Green's functions in neighboring
dimensions. The derivation looks different in even and
odd dimensions, but it leads to the same result, namely (see Appendix \ref{notes})
\begin{equation}
\label{asymptotic3}
\de  h_{MN}\sim \de h_{MN,P}\sim \frac1{r^{(D-2)/2}}.
\end{equation}
In the case of the ADD model the behavior of the field of a uniformly moving mass depends on the ratio of $r$ to
the compactification radius $R$. For $r\ll R$ it has the $D$-dimensional behavior $1/r^{D-3}$,
while for $r\gg R$ it decreases as
$1/r$ (with $r$ being the radial coordinate in three-space). We assume that the observational
wave-zone corresponds to this second condition. Then, the fall-off of the radiation field  in the
wave-zone will be the same as in four dimensions.

Since the  effective stress-tensor $S_{MN}(\de h)$ is a homogeneous second order function of $\de h$ and its
derivatives, its fall-off in $D$-dimensional Minkowski will be $1/r^{(D-2)}$ and in the ADD case
$1/r^2$. This is precisely what is needed to get non-zero flux of radiation in the wave zone through the
$D-2$-dimensional sphere at spatial infinity of $M_{1,D-1}$, or through the corresponding two-dimensional sphere in
the ADD scenario. Note that the use of pseudotensors to define the momentum density of
gravitational radiation leads to the same result. Also, note that as a consequence of the above behavior
the higher order non-linear terms contained in $N_{MN}$ of Eq.(\ref{SMN}) do not contribute to the radiation
energy-momentum.

\subsection{The energy-momentum of the gravitational radiation - Formulae}

Following Weinberg \cite{Weinberg}, we introduce the gravitational energy-momentum
$t_{MN} \equiv G_{MN}-G^{(1)}_{MN}$.
Combined with the matter part $T_{MN}$, the total energy-momentum of the
system is conserved, i.e. it satisfies
\begin{equation}
\label{divweinberg}
\pa^N (T_{MN} + t_{MN})=0\,,
\end{equation}
which is a consequence of the identity $\partial_N G^{(1)MN}=0$, and which to second order in the above
iteration scheme coincides with (\ref{divtau}).

\subsubsection{The uncompactified case $M_{1,D-1}$}

Consider the flat-space world-tube $W$ bounded by two space-like hypersurfaces $\Sigma_{\pm\infty}$
defined at $t\to\pm\infty$, and the infinitely far away time-like Cylindrical surface $C$. Thus, the boundary
$\partial W$ of $W$
is the union of $\Sigma_{-\infty}$, $\Sigma_{+\infty}$ and $C$.

Using the energy-momentum conservation equation
(\ref{divweinberg}) one can write successively for the emitted
radiation momentum $\Delta P_M$ \begin{align} \label{DPM1} \Delta
P_M =&- \int\limits_{\Sigma_{+\infty}} T_{MN} d\sigma^N +
\int\limits_{\Sigma_{-\infty}} T_{MN}
d\sigma^N=-\int\limits_{\partial W} T_{MN} d\sigma^N
=-\int\limits_W
\partial^N T_{MN}d^D x
\nonumber \\
 =&\int\limits_W \partial^N t_{MN} d^D x =\int\limits_{\partial W}
t_{MN} d\sigma^N =\int\limits_C t_{MN} d\sigma^N.
\end{align}
To
justify the above steps, notice that the first equality is (with
$d\sigma^0>0$ in both $\Sigma_{-\infty}$ and $\Sigma_{+\infty}$
surfaces) the statement that $\Delta P_M$ is minus the change of
momentum of the matter system; for the second one uses the fact
that the matter system is localized inside the volume and the
contribution of $C$ on the surface integral vanishes. Gauss'
theorem was used next, together with (\ref{divweinberg}) to end up
with the surface integral of $t_{MN}$. Finally, for the last
equality one used the fact that the surface integrals of $t_{MN}$
on $\Sigma_{-\infty}$ and $\Sigma_{+\infty}$ are zero.

In particular, to the order of our computation in the previous
subsection, using Gauss' theorem, equations (\ref{SMN}),
(\ref{natag_0}) and the gauge fixing condition, one may write
instead of (\ref{DPM1}) \begin{align} \label{DPM2} \Delta
P_M=\int\limits_C S_{MN}(\de h) \, d\sigma^N=\int\limits_W
\partial^N S_{MN}(\de h) \,d^D x =\frac{1}{2} \int\limits_W \de
h_{PQ,M}\, \Box \de\psi^{PQ} d^D x + \int\limits_W \de h^{PQ} \Box
\de\psi_{MP,Q}\, d^D x\,. \end{align}

Furthermore, it is straightforward to show that the last integral
on the right vanishes. Indeed,
\begin{align} \nonumber
\int\limits_W \de h^{PQ} \Box \de\psi_{MP,Q} \,d^D
x&=\int\limits_C \de h^{PQ}\, \Box \de\psi_{MP} \,d\sigma_Q -
\int\limits_W \de h^{PQ}{}_{\!\!,Q}\, \Box \de\psi_{MP} \, d^D x\nonumber \\
&=-\frac{1}{2} \int\limits_W \de h^{,P}\Box \de\psi_{MP}\, d^D x \nonumber \\
&=-\frac{1}{2}\int\limits_C \de h\,\Box \de\psi_{MP}\, d\sigma^P +
\frac{1}{2} \int\limits_W \de h\,\Box \de\psi_{MP}{}^{\!,P}\, d^D
x=0\,,
\end{align} where the first integrals of the first and
third lines vanish as a consequence of the asymptotic behavior of
the fields given in (\ref{asymptotic1}), (\ref{asymptotic2}), and
(\ref{asymptotic3}).

Thus, the emitted momentum is
\begin{equation}
\label{DPMM1} \Delta P_M=\frac{1}{2} \int\limits_W \de h_{PQ,M}
\Box \de\psi^{PQ} \, d^Dx\,. \end{equation}

Using the Fourier transformed quantities
\begin{align}
\label{gr_pert9} \de h^{MN}(x)=\frac{1}{(2 \pi)^D}\int  \de h^{MN}
(k) \,e^{-i k_Q x^Q} d^D k\, , \qquad \tau^{MN}(x)=\frac{1}{(2
\pi)^D}\int \tau^{MN}(k) \,e^{-i k_Q x^Q} d^D k.
 \end{align}
one rewrites (\ref{DPMM1}) in the form
\begin{align}
\label{gr_pert11}
\Delta P_M =-\frac{i\varkappa_D^2}{2(2 \pi)^D} \int
k_{M}G_{\rm ret}(k) \tau_{SN}(k)
\tau^*_{LR}(k)\tilde{\Lambda}^{SNLR}\, d^D k,
\end{align}
where $\tau_{MN}^*(k)=\tau_{MN}(-k)$ by the reality of $\tau_{MN}(x)$, and
the tensor $\tilde\Lambda^{SNLR}$ is given by
\begin{align}
\label{gr perts11a}
\tilde{\Lambda}^{SNLR}=\frac{1}{2}\left[\eta^{SL} \eta^{
NR}+\eta^{SR} \eta^{NL}\right]-\frac{1}{D-2}\eta^{SN} \eta^{ LR}.
\end{align}

The retarded Green's function is $G_{\rm ret}=-{\mathcal P}(1/k^2)+i\pi\epsilon(k^0)\delta(k^2)$. Its real part leads
to an integrand, which is odd under parity $k^M\to -k^M$ and does not contribute to the integral.
Thus, one writes equivalently
\begin{align}
\label{gr_pert13}
\Delta P_M=\frac{\varkappa_D^2}{2 (2 \pi)^{D-1}}
 \int \theta(k^0)
 k_{M}  \tau_{ SN}(k)
\tau^*_{LR}(k)\tilde\Lambda^{SNLR} \delta(k^2) \, d^D k \,.
\end{align}
Taking into account the transversality of $\tau_{MN}$ ($k^M\tau_{MN}(k)=0$) and the on-shell condition
$k^2=0$ of the emitted wave, one can replace the Minkowski metric in
$\tilde{\Lambda}^{SNLR}$ by
\begin{align}
\label{gdddd}
\Delta^{MN} \equiv ~_{g}\Pi^{M}_{ \, \,L} ~_{k'}\Pi^{LN
}=\eta^{MN}+\frac{k^{M}k^{N}-2 (k g)k^{(\!M}g^{N)}}{(kg)^2} \, ,
\end{align}
with any time-like unit vector $g$, where $~_{g}\Pi=1-g\otimes g$ and $~_{k'}\Pi=1+\,
k'\otimes k'/(kg)^2$ are projectors onto subspaces transverse to $g$ and
$k'\equiv~_{g}\Pi k = k-(kg)g$, respectively. Since $(k'g)=0$,
the projectors $ ~_{g}\Pi$ and $~_{k'}\Pi$ commute. Their product
$\Delta^{MN}$ is then a symmetric  projector onto the subspace
$M_{k,g}$, perpendicular to $k$ and  $g.$ By
construction, the projector $\Delta$ is idempotent ($\Delta^2=\Delta$), thus on $M_{k,g}$ it acts as the unit
operator. In what follows, we will conveniently choose $g_M=u'_M$ and calculate the flux in the Lorentz frame
(the Lab) with $u'_M=(1,0, \ldots, 0)$.

Thus, $\tilde\Lambda$ can equivalently be replaced in (\ref{gr_pert13}) by
\begin{align}
\label{gr perts11b}
 {\Lambda}^{SNLR}=\frac{1}{2}\left[\Delta^{SL} \Delta^{ NR}+\Delta^{SR}
\Delta^{NL}\right]-\frac{1}{D-2}\Delta^{SN}\Delta^{ LR} ,
\end{align}
and then upon integration over $|\mathbf{k}|$, one ends up with
\begin{align}
\label{gr_pert14}
\Delta P_M=\frac{\varkappa_D^2}{4(2 \pi)^{D-1}}
\int\limits_{0}^{\infty}\omega^{D-3} d\omega \int\limits_{S^{D-2}}
d\Omega \;k_{M}  \tau_{ SN}(k)
\tau^*_{LR}(k) \Lambda^{SNLR}\,,
\end{align}
where $\om\equiv k^0=|{\bf k}|$.
It is easy to show that $\Delta\equiv \eta_{MN} \Delta^{MN}=D-2$. Furthermore, $\Lambda$ satisfies the
following relations:

\noindent
(a) It is traceless on both pairs of indices
\begin{align}
\label{gr perts11c}
\eta_{SN}\,{\Lambda}^{SNLR}=0 \,, \quad  \eta_{LR} \,{\Lambda}^{SNLR}=0\,,
\end{align}
and (b) it is a projection operator, since it is idempotent  ($\Lambda^2=\Lambda$)
\begin{align}
\label{idempLambda}
\eta_{PS}\eta_{QT}\Lambda^{MNST}  \Lambda^{PQLR}  =\Lambda^{MNLR}\,,
\end{align}
and projects any second rank symmetric tensor, element of a space
we denote by $K$, onto a subspace $K_\Lambda$. Acting on a
$\tau_{MN}$ (or $\tau_{LR}$) \footnote{The arguments below apply
independently to the real and the imaginary parts of the
complex-valued $\tau_{MN}$ that enters in (\ref{gr_pert14}). }, it
returns a new symmetric tensor ${\tau'}^{MN}=\Lambda^{MNPQ}
\tau_{PQ}$ with two indices. $\Lambda$ acts as a linear operator
on the space $K$ of symmetric tensors of rank two \footnote{This
becomes even more evident if one denotes the pair of indices
$(MN)$ as $a$ and the pair $(LR)$ as $b$. Then the action of
$\Lambda$ on $\tau$ takes the form
${\tau'}^a=\Lambda^{ab}\tau_b$.}. The properties of $\tau'$
determine the space $K_\Lambda$.

In particular, the dimensionality of $K_\Lambda$ is determined as
follows: $\tau'_{MN}$ is a symmetric $D\times D$ matrix and thus
it has $D(D+1)/2$ elements. In addition it satisfies the
independent conditions $\tau'_{MN} {u'}^{N}=0$, $\tau'_{MN}
k^{N}=0$ and $\eta^{MN} \tau'_{MN}=0$, which  impose \footnote{The
condition $\tau'_{MN} k^{N}=0$ gives $D-1$ constraints. This is
due to the fact that $k^2=0$, which allows an arbitrariness in
$\tau'$ of the form $\alpha k_M k_N$.} $D+(D-1)+1=2D$ conditions,
leaving a total of $D(D-3)/2$ independent elements in $\tau'$.
Thus, the dimensionality of $K_\Lambda$ is $D(D-3)/2$, the same as
the number of independent polarizations of the graviton in
$D-$dimensions.

Given that $\Lambda$ acts on $K_\Lambda$ as the unit operator one can write it (decomposition of unity) in terms
of an orthonormal basis tensors \{$\varepsilon_{\mathcal P}^{MN}$\}, with ${\mathcal P}=1,2,\ldots,D(D-3)/2$
of $K_\Lambda$
\begin{align}
\label{gr perts10001}
\Lambda^{MNLR}=\sum_{\cp} \varepsilon^{MN}_{\cp}\varepsilon^{LR}_{\cp}\,.
\end{align}
Using this expression for $\Lambda$, equation (\ref{gr_pert14}) takes the form
\begin{align}
\label{DEMD}
\Delta P_{M}= \frac{\varkappa_D^2}{4(2 \pi)^{D-1}}
\sum_{\cp } \int\limits_{0}^{\infty}\od^{D-3} d\od \!\!
\int\limits_{S^{D-2}} \!\! d\Omega \;
 k_{M} \left|
\tau_{LR}(k)\,\varepsilon^{LR}_{\cp}\right|^2 \,,
\end{align}
i.e. a sum of independent contributions one for each polarization.

\subsubsection{The compactified case $M_{1,3}\times T^d$}

Following the same steps as in $M_{1,D-1}$, one is led to a similar expression for the emitted
momentum in the ADD scenario. Consider the flat-space world-tube $W$ bounded by two space-like
hypersurfaces $\Sigma_{\pm\infty}$ defined at $t\to\pm\infty$, and
the infinitely distant time-like surface $S_{r\to\infty}$, defined
for $-\infty<t<\infty$, and whose intersection with the space-like hypersurfaces
$t={\rm const}$ will be denoted by $B_t$. According to our convention all
quantities constructed from the metric perturbation $h_{MN}$ are
flat-space tensors, so the covariant derivatives acting on them are simply partial derivatives.
Starting with the energy-momentum conservation (\ref{divweinberg}), and following the same steps as above,
one concludes, first of all, that the change of the momentum of the system carried away by
gravitational waves during the collision can be computed by integrating
the flux of momentum through $B_t$ and also over time
 \begin{equation}
 \Delta P^{M}=\int\limits_{-\infty}^{\infty}dt \int\limits_{B_t}
 S^{MN}\lbr h \rbr d^{D-2}B_N.
 \label{DPMadd1}
 \end{equation}
Note that the hypersurface $B_t$ is the product $S^2\times T^d$ and the
corresponding measure is $d^{D-2}B_N=r^2 d\Omega_2 \,d^d\y$, with $r$ the radial coordinate in
$R^3$. Thus, using Gauss' theorem one may write:
\begin{equation}
\int\limits_W \pa_N S^{MN}d^4x \,d^d\y=
 \int\limits_{\Sigma_{\infty}} S^{M0}d^3x\, d^d\y -\int\limits_{\Sigma_{ -\infty}} S^{M0}d^3x\, d^d\y
+\lim_{r\to \infty}\int\limits_{-\infty}^{\infty}dt \int\limits_{T^d}d^d\y\oint
S^{Mr} r^{2}d\Omega_{2}.
 \end{equation}
 Given that for $t=\pm \infty$ the metric is flat
 \footnote{In this calculation we
ignore the self-action of  fields upon the particles, but even if
we did not, the change of (infinite) self-field momenta would be
zero.}, the contribution of the integrals over
$\Sigma_{\pm\infty}$ is zero. Taking, in addition, into account
the asymptotic behavior of $\de h$ and its derivatives (see
Appendix \ref{notes}), one concludes that (\ref{DPMadd1}) takes
the form:
 \begin{equation}
 \Delta P^{M}=\int\limits_W
\pa_N S^{MN}\lbr\de h\rbr d^Dx\,.
\end{equation}
Finally, repeating the same steps and arguments as between equations (\ref{DPM2}) and (\ref{DPMM1}) and
taking into account the asymptotic behavior relevant to the case at hand,
one ends up with the same expression as in the $M_{1,D-1}$ case for the emitted momentum
\begin{equation}
\label{losfinal} \Delta P_M=\fr{1}2 \int\limits_W \de  h_{PQ,M} \,
\Box \, \de \psi^{PQ}\;d^Dx\,.
\end{equation}

In terms of the Fourier transformed quantities defined by
 \begin{align}
 \label{furADD}
 & h^{MN}(x,y)=\frac{1}{(2 \pi)^4 V}\sum_{n}\int  h_{(n)}^{MN}
(k^\mu)\, e^{-i k_\mu x^\mu+i n_i y^i/R} d^4 k\, , \nn \\
& \tau^{MN}(x,y)=\frac{1}{(2 \pi)^4 V}\sum_{n}\int \tau^{MN}_{(n)}
(k^\mu)\, e^{-i k_\mu x^\mu+i n_i y^i/R} d^4 k\, .
\end{align}
($V$ is the volume of the $d-$dimensional torus) the emitted momentum in the transverse to the brane
directions takes the form
\begin{equation}
\label{Mom5} \Delta P^i=\int\limits_{\mathbb{R}^3\times T^d}  \!\!
\tau^{i0} d^3 x \,d^d\y \sim \int dk^0 dk'^0 \sum_{n \in
\mathbb{Z}^d}\frac{\e^{i(k^0-k'^0)t}k^0 }{[(k^0)^2-{\bf
k}^2-\uln^2]
 [(k'^0)^2-{\bf k}^2-\uln^2]}  {n}^i,
\end{equation}
where ${\bf k}$ is a wave 3-vector on the brane and $k_T^2 \equiv
 {n}^2 /R^2$. $\Delta P^i$ vanishes since the integrand is odd under parity ${\bf n}\to {\bf -n}$.
The emitted momentum tangential to the brane can be obtained along
the same lines as above with the only difference of using the
discrete Fourier transformation on the torus. Finally, the emitted
energy ($E \equiv \Delta P_0$) is
\begin{align}
\label{gr_pert11ADD}
  E=\frac{\varkappa_D^2}{32 \pi^{3} V}\sum_{n \in
\mathbb{Z}^d} \int\limits_{0}^{\infty}\!\mathbf{k}^{2}\,
d|\mathbf{k}| \int\limits_{S^{2}}\! d\Omega  \tau_{SN}(k)\,
\tau^*_{LR}(k)\, {\Lambda}^{SNLR} \left. \vphantom{\sqrt{d}}
\right|_{k^0=\sqrt{{\bf k}^2+\uln^2}},
\end{align}
where  $\tau_{MN}(k)$ is the four-dimensional Fourier-transform of
the total source evaluated at $k^0=\sqrt{\mathbf{k}^2+\uln^2}$.

\begin{figure}
\begin{center}
\includegraphics[angle=0,width=15cm]{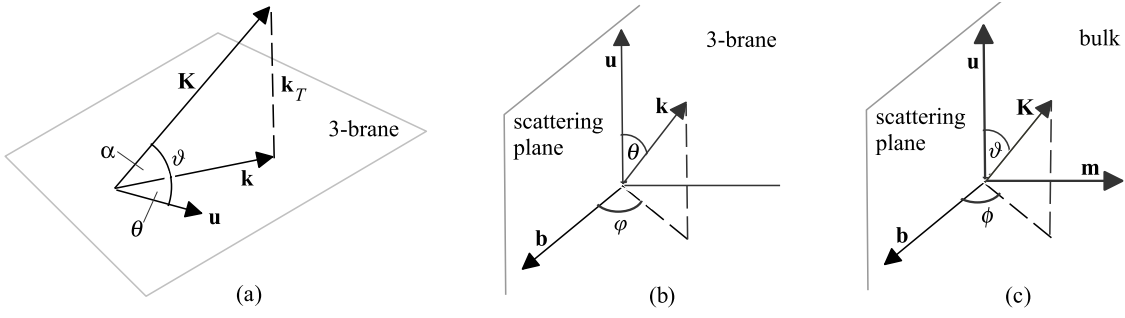}
\caption{The angles in lab frame used in the text.}
\label{branepic}
\end{center}
\end{figure}

The wave vector of the emitted gravitational wave is a genuine $D$-dimensional vector with
quantized transverse components. We denote
\begin{equation}
\label{kam} k^M=(k^\mu,\; \uln^i), \qquad \uln^i=\frac{n^i}{R}\,,
\qquad n^i\in \mathbb{Z}.
\end{equation}
The notation and conventions for the angles used in this paper is shown in Figure \ref{branepic}.
For the brane four-vector $k^\mu$ we use the following parametrization ensuring zero square of $k^M$:
\begin{equation}
\label{kamu} k^\mu=\left(\sqrt{\varpi^2+k_T^2},\;
 \varpi \sin\theta\cos\ffi,\; \varpi
\sin\theta\sin\ffi,\;\varpi\cos\theta\right)\, .
\end{equation}

Finally, using (\ref{gr perts10001}) the formula for the emitted energy becomes
\begin{equation}
\label{DEadd}   E=\frac{\varkappa_D^2}{32 \pi^{3} V}\sum_{\mathcal
P} \sum_{n\in \mathbb{Z}^d} \int\limits_{0}^{\infty}\!\!
\varpi^{2}\, d \varpi \int\limits_{S^{2}}\!\! d\Omega\, \left
|\tau_{SN}(k)\,\varepsilon^{SN}_{\mathcal P}\right|^2 \left.
\vphantom{\sqrt{d}} \right|_{k^0=\sqrt{ \varpi^2+\uln^2}}\,.
\end{equation}

\subsection{Graviton polarizations}

The computation of the emitted energy can be simplified considerably if one chooses appropriately
the orthonormal basis of the $D(D-3)/2$ graviton polarization tensors \{$\varepsilon^{\mathcal P}_{MN}$\}
that appear in (\ref{DEMD}) and (\ref{DEadd}). The construction of such a convenient basis is presented here.

\vspace{0.3cm}
$\bullet\;${\it The first step} towards their construction is to define  $D-2$ space-like unit vectors orthogonal
to $k$ and $u'$ and among themselves. The first $D-4$  vectors $e_\al^M,\,
\al=3,\ldots,D-2$ we choose to be orthogonal to the collision
three-dimensional subspace spanned by the vectors
$u^M,\,u'^M,\,b^M$ which lie on the brane and together with $k$ fully characterize the kinematics of the problem
under study. Thus, they are taken to satisfy  the orthogonality properties
\begin{align}
\label{xxx1s3} \quad e_{\alpha}^M e_{\beta M}=-\delta_{\alpha
\beta},\qquad (e_{\alpha}
k)=(e_{\alpha}u')=(e_{\alpha}u)=(e_{\alpha} b)=0\,.
\end{align}
As mentioned above, in the lab frame $u'^M=(1, 0, \pp, 0)$. Also, with no loss of generality the vectors
$e_{\alpha}$ can be chosen to have vanishing components along the brane.

Consider next that the brane Cartesian coordinate frame
$\{t,x,y,z\}$ together with the above $D-4$ space-like unit
vectors $e_{\alpha}^M $ form a $D$-dimensional {\it right-handed}
basis $\{t,x,y,z,e_3,\ldots, e_{D-2}\}$ and choose the
$D$-dimensional Levi-Civita symbol so that $\epsilon^{0xyz3\ldots
(D-2)}=1$. The two remaining space-like vectors $e_1$ and $e_2$
are defined as follows: $e_2^M$ is chosen orthogonal to both
$u^M,\,u'^M$, namely
\begin{equation}\label{vector_e2}
e_2^M=N^{-1}\epsilon^{M M_1 M_2 \ldots
M_{D-1}}u_{M_1}u'_{M_2}k_{M_3}e_{3 M_4}\ldots e_{(D-2)\,M_{D-1}}\,
,
\end{equation}
where $N$ is the normalization factor which can be obtained by squaring the above expression.
Specifically, making use of the relation $k^2=0$ one obtains
\begin{equation}
N^2=-\left[(u'k)u-(uk)u'\right]^2.
\end{equation}
According to this definition
\begin{equation}\label{e2products}
(e_2 u)=(e_2 u')=(e_2 k)=(e_2 e_\al)=0\,, \quad \alpha=3,4,\ldots, D-2\,.
\end{equation}
Finally, the remaining vector $e_1$ is chosen orthogonal to $u'$ and to all previously constructed
polarization vectors. It is given by
\begin{equation}\label{vector_e1}
e_1^M=N^{-1}\!
\left[(ku)\,u'^M-(ku')\,u^M+\left((uu')-\fr{(ku)}{(ku')}\right)k^M\right]\,
,
\end{equation}
with the same normalization factor. Note that $e_2^M$ is
confined on the brane, while $e_1^M$ has both brane and bulk
components due to $k^M$.

\vspace{0.3cm}

$\bullet\;${\it The next step} is to use the above basis of orthonormal polarization vectors to build the
$N_{\cp}=D(D-3)/2$ traceless mutually orthogonal polarization tensors $\ep_{\cp}^{MN}$, i.e. satisfying:
\begin{equation}
\label{epcond}
\ep^{MN}_{\cp}\eta_{MN}=0\;\; \;{\text{and}} \;\;\; \ep_{\cp}^{MN}
\ep_{\cp'\,MN}=\delta_{\cp \cp'}\,.
\end{equation}
First, we build from $e_1$ and $e_2$ two tensors, the ones familiar from four-dimensional gravity:
\begin{equation}
\ep_{(\mathrm{I})}^{MN}=\frac{e_1^{M}e_1^{N}-e_2^{M}e_2^{N}}{\sqrt{2}},\qquad
\ep_{(\mathrm{II})}^{MN}=\frac{e_1^{M}e_2^{N}+
e_2^{M}e_1^{N}}{\sqrt{2}},
\end{equation}
or, equivalently, the corresponding chiral ones
\begin{equation}
\ep_{\pm}^{MN}=\frac{\varepsilon^{MN}_{(\mathrm{I})}\pm
i\ep^{MN}_{(\mathrm{II})}}{\sqrt{2}}=e_{\pm}^{M}e_{\pm}^{N},\qquad
e_{\pm}^{M}=\fr{e_1^{M}\pm i e_2^{M}}{\sqrt{2}}.
\end{equation}
The remaining set of tensors is taken to have the same general form as the type-I and type-II above,
namely to have (up to coefficients) the form of a sum of same index binomials $e_i^{M} e_i^{N}$ for the first kind,
and different index binomials $e_i^{(M} e_j^{N)}$ for tensors of the 2nd type.
Thus, the third tensor involving $e_1,\,e_2$, contains the diagonal sum of the $D-4$ vectors $e_\al^M$
\begin{equation}
\varepsilon_{(\mathrm{III})}^{MN}=\sqrt{\frac{2(D-4)}{D-2}}
\left[\frac{1}{D-4}\sum_{\alpha=3}^{D-2}
e_\alpha^{M}e_\alpha^{N}-\frac{e_1^{M}e_1^{N}+e_2^{M}e_2^{N}}{2}\right].
\end{equation}
For $D > 4$ one can build the larger set of $(D-2)(D-3)/2$
tensors similar to $\ep_{(\mathrm{II})}$, starting with the full set
of unit vectors $e_1,\;e_2,\;e_\al$, namely
\begin{align}\label{aaa2}
\varepsilon_{(ij)}^{MN}=\frac{e_i^{M}e_j^{N}+e_j^{M}e_i^{N}}{\sqrt{2}},\qquad
i< j, \qquad j=1,2,\ldots, D-2.
\end{align}
Finally, the remaining $D-5$ polarization tensors $\varepsilon_\alpha$ are obtained from the linearly
independent tensors
\begin{align}
\label{aaa4}
f_{\al}^{MN}=\frac{1}{\sqrt{(D-5)(D-4)}}\left[(D-5)e_\al^{M}e_\al^{N}-\sum_{\beta
\neq \al} e_\beta^{M}e_\beta^{N}\right], \qquad \al=3, \ldots, D-3.
\end{align}
built from $e_\al$, after ortho-normalization to satisfy the
conditions (\ref{epcond}) \footnote{One could a priori take
$\alpha=3, 4, \ldots, D-2$ for the label of the polarizations
$f_\alpha$. However,  they are not independent, since as one can
easily verify $\displaystyle \sum_{\alpha=3} ^{D-2}
f_\alpha^{MN}=0$.}. For $D\leqslant 5$ there are no such
polarizations.

\vspace{0.3cm}
$\bullet$ {\it Finally}, one can verify that $\Lambda^{MNLR}$ can be written in terms of these $D(D-3)/2$
polarization tensors $\ep_{\cp},\;{\cp}\mathrm{=I, II, III,} (ij), \al$, in the form
\begin{align}
\label{gr perts101}
\Lambda^{MNLR}=\sum_{\cp}^{N_\cp}\varepsilon^{MN}_{\cp}\varepsilon^{LR}_{\cp}\,.
\end{align}
As advertised, the advantage of using the above special choice of
polarization tensors in (\ref{DEMD}) and (\ref{DEadd}) is now
evident. The indices of the energy-momentum tensor $\tau_{MN}$ are
carried by the four independent vectors $u, u', b, k$, of
which the first three belong to the brane, while $k^M$ is
multidimensional. These four entirely characterize the
bremsstrahlung process. A consequence of the fact that the vectors
$e_\al$ are by construction orthogonal to those, is that only the
three polarizations, namely with $\cp=\pm$ and III contribute to
(\ref{DEMD}) and (\ref{DEadd}). Moreover, the $e_\al$ terms in
$e_{\rm III}$ can also be dropped and the latter simplifies to the
``effective" polarisation:
\begin{equation}
\varepsilon_{(\mathrm{III})}^{MN}=-\sqrt{\frac{D-4}{2(D-2)}}
\left[e_1^{M}e_1^{N}+e_2^{M}e_2^{N}\right]=-\sqrt{\frac{2(D-4)}{D-2}}\,
e_+^{(M}e_{-}^{N)}\,.
\end{equation}
Incidentally, notice that this effective
$\varepsilon_{(\mathrm{III})}^{MN}$ vanishes, as it should, in the
special case of four dimensions.

Having chosen the vectors $e_1^M$ and $e_2^M$ to be orthogonal to $k^M$ and $u'^{M}$,
only the scalar products of $e_1^M$ and $e_2^M$ with $u^M$ and $b^M$ will enter in the computation
of the radiated energy momentum. According to
(\ref{e2products}), $(e_2 u)$ vanishes. The remaining
non-vanishing products $(e_1 b)$, $(e_1 u)$ and $(e_2 b)$ are
given in the Appendix \ref{formulae}.

\section{The gravitational radiation source}
\label{gr_gr}

To linear order the metric perturbation is just the sum
$h_{MN}+h'_{MN}$ of the separate contributions of the two
particles. They are for each particle the solution of
(\ref{hanuleq}) with the corresponding zeroth order energy
momentum tensor (\ref{T0mn}) for its source. In Fourier space they
are given by \footnote{It is worth noting that in four dimensions
and in the center of mass frame the limit of $h_{MN}$ for $m\to 0$
and $\gamma\to \infty$ with $\mathcal{E}=m\gamma$ fixed, is the
Aichelburg-Sexl metric \cite{Aichelburg}.}
 \begin{align}
 \label{grsola}
 &h_{MN}=\sum_{l \in \mathbb{Z}^d} h^{l}_{MN}(q)\,, \qquad  h^{l}_{MN}(q)=
 \frac{2\pi \varkappa_D m}{q^2-q_T^2}\e^{iqb}\delta(qu)\left( u_{M}u_{N} -
\frac{1}{d+2}\eta_{MN} \right);   \\
&h'_{MN}=\sum_{l \in
\mathbb{Z}^d} {h'}^{l}_{MN}(q)\,,  \qquad {h'}^{l}_{MN}(q)=
 \frac{2\pi \varkappa_D m'}{q^2-q_T^2}\delta(qu')\left(u'_{M}u'_{N} -
\frac{1}{d+2}\eta_{MN} \right)\,,
\label{1h'mn}
\end{align}
where $q^i_T=l^i/R$, the transverse components of the momentum,
and $ q_T^2=\sum (q^i_T)^2$.

The deflection of the trajectory of particle $m$ due to its gravitational interaction with $m'$ is
obtained by inserting the above $h'_{MN}$ in (\ref{z1eq}) and solving for $\un z^M$. One obtains
\begin{align}
\label{sc6}
\un z^{M}(s)=-\frac{ i m' \varkappa_D^2}{(2
\pi)^{3} V}  \sum_{l \in \mathbb{Z}^d} \int d^{4} q
\frac{\delta(qu')}{(q^2-q_T^2)
(qu)}e^{-iqb}\left(e^{-i(qu)s}-1\right) \left[\gamma
u'^{M}-\frac{1}{d+2}u^{M}-\frac{ \gamma_*^2}{2(qu)}
q^{M}\right]\, ,
\end{align}
where $\gamma_*^2 \equiv \gamma^2-1/(d+2)$. Similarly
for $\un {z'}^M$. As expected the trajectory deviation in the
transverse directions
\begin{align}
\label{ygrek}
\delta y^i \sim \sum_{l \in \mathbb{Z}^d}\int d^4 q
\frac{q^i_T}{q^2-q_T^2}\frac{\delta(qu')}{(qu)^2}e^{-iqb}\left(e^{-i(qu)s}-1\right)
\end{align}
vanishes, since the summand is odd under parity ($q_T^i \to -q_T^i$), and as expected the particles do not
leave the scattering plane.

\subsection{The local part $\un T_{MN}+\un T'_{MN}$ of the source}

The contributions of the particle trajectories to the gravitational radiation source is given in (\ref{T1MN}).
They can equivalently be written in the form
\begin{align}\label{gg1}
\un T_{MN}(k)=e^{ikz(0)} m \int d\tau \, e^{i(ku)\tau}\left[ i(k\un
z)\, u_{M}u_{N}+2 u_{(M} \un \dot{z}_{N)}-\vk \frac{h'}{2}u_{M}u_{N}+2 \vk u^{S}h'_{S
(M}u_{N)}\right]\, ,
\end{align}
and
\begin{align}\label{gg1st}
\un T'_{MN}(k)= e^{ikz'(0)} m' \int d\tau \, e^{i(ku')\tau}\left[
i(k\un z')\,u'_{M}u'_{N}+2 u'_{(M} \un
\dot{z}'_{N)}-\vk \frac{h}{2}u'_{M}u'_{N}+2\vk u'_{S}h^{S}_{\;
(M}u'_{N)}\right] \,,
\end{align}
respectively.
With the specific choice of polarization tensors made above, the projection of
$\un T'_{MN}(k)$ on any of them vanishes. Then, substituting the $h'$ and $\un z$ given in
(\ref{1h'mn}) and (\ref{sc6}), one obtains:
\footnote{Note, incidentally, that as expected the part of $\un z_M$ which corresponds to uniform motion, i.e. the
term $-1$ in the first parenthesis, does not contribute to (\ref{gg1_5}).
}
\begin{align}
\label{gg1_5}
&\un T_{MN}(k)=\frac{mm' \varkappa^2}{(2\pi)^{2}V} \,e^{ikb}
\sum_{l \in \mathbb{Z}^d} \left[\left( \gamma\frac{ku'}{ku}I^l -
\frac{\gamma_*^2}{2(ku)^2}(k \cdot I^l)
 \right)u_{M}u_{N}+ \frac{\gamma_*^2}{(ku)} u_{(M}I^l_{N)}\right]\,,
\end{align}
in terms of the integrals $I^l$ and $I^l_M$, defined by
\begin{align}
I^{l} = \int \frac{ \delta(pu')\,
\delta(ku-pu)\,\e^{-i(pb)}}{p^2-\ull^2}\;d^4 p\,,      \quad
I_{M}^{l}
= \int \frac{ \delta(pu')\, \delta(ku-pu)\, \e^{- i(pb)}}{p^2-\ull^2} \;p_{M} \;  d^4 p\,,
\end{align}
with $p_T^i=l^i/R$. These were computed in \cite{GKST-1} and
expressed in terms of Macdonald functions
\begin{align}
I^{l}  =-\frac{2\pi}{\gamma v} {K}_{0}(z_l)\,, \quad
I_{M}^{l}
 =-\frac{2\pi }{\gamma v b^{2}}\left(b z {K}_{0}(z_{l})\frac{\gamma
u'_{M}-u_{M}}{\gamma v}+ i \hat {K}_{1}(z_{l})\,
b_{M}\right),
\end{align}
with argument
\begin{equation}
\label{z1}
 z_{l}\equiv (z^2+\ull^2 b^2)^{1/2}\,,  \qquad z\equiv \fr{(ku) b}{\gamma v}\, ,
\end{equation}
while the hatted Macdonald functions are defined by
$\hat{K}_{\nu}(x)\equiv x^{\nu}{K}_{\nu}(x)$.

Substituting back into (\ref{gg1_5}) one obtains
\begin{align}
\label{gg2}
\un T_{MN}(k)=-\frac{m m' \varkappa^2}{4\pi
V}\frac{e^{ikb}}{\gamma v^3} \!\sum_{l \in \mathbb{Z}^d} \!\left[-i
\Gamma \frac{\hat{K}_{1}(z_l)}{z^2}\sigma_{MN} - 2
\Gamma\gamma  {K}_{0}(z_l) \, {u\bl}_{(M}{u'}_{N)} \!
   + \!\left(\! \left(2v^2 \!- \! \Gamma\right)\gamma \frac{z'}{z} -\Gamma\! \right) {K}_{0}(z_l) \, u_{M}u_{N} \right] ,
\end{align}
where
\begin{align}
\label{gg3}
\sigma_{MN}\equiv (kb) u_{M}u_{N} -2(ku)u_{(M}b_{N)} \quad {\rm and} \quad
\Gamma \equiv \frac{\gamma_\ast^2}{\gamma^2}= 1-\frac{1}{
(d+2)\gamma^2}\,.
\end{align}
Finally, omitting the terms longitudinal in $u'$, since they give zero when contracted with any
polarization tensor, the above expression simplifies to
\footnote{In what follows we will omit from the radiation source all terms,
which give zero when contracted with all polarization tensors, since they do not contribute to the emitted
energy-momentum.
}
\begin{align}
\label{gg2_5}
 \un T_{MN}(k)=-\frac{m m' \varkappa^2}{4\pi
V}\frac{e^{ikb}}{\gamma v^3} \!\sum_{l \in \mathbb{Z}^d} \!\left[-i
\Gamma \frac{\hat{K}_{1}(z_l)}{z^2}\sigma_{MN}
   + \!\left(\! \left(2v^2 \!- \! \Gamma\right)\gamma \frac{z'}{z} -\Gamma\! \right) {K}_{0}(z_l) \, u_{M}u_{N} \right]\,.
\end{align}
As in previous analogous cases \cite{GKST-1, GKST-3}, the
exponential fall-off of the Macdonald functions for large values
of the argument $z_l$ leads to an effective natural cut-off
$N_{\rm int}$ on the number of interaction modes $l$ in the sum.
One can estimate the radius $l_{\rm int}$ of the sphere in the
$\{l^i\}$ space, beyond which the modes can be neglected, by
setting
\begin{equation}
\label{lmax}
{\left(l_{\rm int}/R\right)}^2\, b^2\sim 1 \,,
\end{equation}
from which the number of contributing interaction modes is obtained
\begin{equation}
N_{\rm int}\sim \left(\frac{R}{b}\right)^d \sim \frac{V}{b^{\,d}}\,.
\end{equation}
For $b\ll R$, which is the case of interest here, one has $N_{\rm
int} \gg 1$, and the mode-summation can be converted to
integration using \cite{GKST-1} ($Z>0$)
\begin{align}
\label{sum2int} \frac{1}{V} \sum_{l}\hat{K}_{
\lambda}\left(\sqrt{Z^2+{\ull}^2 b^2}\right) \simeq
\frac{1}{(2\pi )^{d/2} b^d} \hat{K}_{\lambda+d/2}(Z)\,,
\end{align}
to end up with
\begin{align}
\label{Tloc}
\un T_{MN}(k)=- \frac{ \lambda \, e^{i(kb)}}{\gamma v^3} \left[-i
\Gamma \frac{\hat{K}_{d/2+1}(z)}{z^2}\sigma_{MN}
+ \!\left(\! \left(2v^2 \!- \! \Gamma\right)\gamma \frac{z'}{z} -\Gamma\! \right) \hat{K}_{d/2}(z) \, u_{M}u_{N} \right]\,,
\end{align}
where
\begin{equation}
\label{lambda}
\lambda\equiv \frac{\vk^2 m' m}{2 (2\pi)^{d/2+1} b^d}\,.
\end{equation}

{\it Angular and frequency characteristics.} The local radiation
source (\ref{Tloc}) above is expressed solely in terms of
Macdonald functions with argument $z$. As has been demonstrated in
analogous instances before \cite{GKST-1, GKST-3}, the exponential
fall-off of these functions implies an effective cut-off of
${\mathcal O}(1)$ in the angle and frequency dependent variable
$z$, which eventually constrains the angular and frequency
distribution of the radiation. Below, it will be shown that a
similar situation arises as well with the non-local amplitude
$S_{MN}$.

\subsection{The non-local part $S_{MN}$ of the source}

The stress tensor $S_{MN}$ is defined in (\ref{natag}) and
calculated in coordinate representation in (\ref{natag_0}). Again,
since its trace does not contribute to the emitted momentum, one
may restrict himself to its traceless part. Write, as explained above,
$h_{MN}+h'_{MN}$ for the leading metric perturbation and
substitute in $S_{MN}$, while omitting self-action terms, to
obtain for its $n$-th mode (in analogy to (\ref{furADD}))
\begin{align}
 \lb{son}
 S_{MN}^{(n)}(x^{\mu})= \frac{1}{V}\sum_{l} & \left[
{h}_{M}^{(n-l)A \cd B}\left({h'}_{NB\cd
A}^{(l)} - {h'}_{NA\cd
B}^{(l)}\right)-\frac{1}{2}h_{AB \cd N}^{(n-l)}
{h'}^{(l)AB}_{\qquad \cd M} -
\frac{1}{2}h_{MN}^{(n-l)}\Box  {h'}^{(l)}+\right. \nn
\\  & \left. \; +\left(h_{MA \cd
NB}^{(n-l)}+ h_{NA \cd MB}^{(n-l)}- h_{AB \cd MN}^{(n-l)}- h_{MN
\cd AB}^{(n-l)}\right){h'}^{(l)AB}
\vphantom{\frac{1}{2}}\right]+\{ h \longleftrightarrow h' \}\, ,
\end{align}
and with $S_{MN}^{(n)}(x^{\mu})$ the four-dimensional
Fourier-transform of $S_{MN}^{(n)}(k^{\mu})$. Upon substitution of
the solutions (\ref{grsola}) we have in momentum representation
\footnote{All terms, which upon contraction with the polarization
tensors give zero, have been omitted from $S^{(n)}_{MN}(k)$.}
\begin{align}
\label{Smn0}
 S_{MN}^{(n)}(k)= \frac{m m' \vk^2 e^{i(kb)} }{4\pi^2 }
   \left[ u_M u_N (ku')^2 J^n(k)+2\gamma(ku') u^{\phantom{n}}_{(M} J^n_{N)}(k)  +\gamma_{*}^2 J^n_{MN}(k) \right]  \,.
\end{align}
The non-vanishing components of the scalar, vector and tensor integrals that appear in (\ref{Smn0}),
namely
\begin{align}
J^n(k)\equiv \frac{1}{V} \sum_{l } \int  \,d^4 p
\frac{\delta(pu')\delta(ku-pu) \, e^{-i(pb)}}{ (p^2-\ull^2)\,
[(k-p)^2 - (\uln^i - \ull^i)^2]}\,, \qquad p^{M}=(p^{\mu},
p_T^i)\,, \qquad p^2 \equiv p_{\mu}p^{\nu}
\end{align}
\begin{align}
J^n_{M}(k)\equiv \frac{1}{V} \sum_{l } \int \,d^4 p \,
\frac{\delta(pu')\delta(ku-pu)\, e^{-i(pb)}}{ (p^2-\ull^2)\, [
(k-p)^2 - (\uln^i - \ull^i)^2] }\, p_{M} \,,
\end{align}
and\footnote{Note that the bulk components $J^n_{i}$ and
$J^n_{iM}$ vanish by parity only if $k_T^i=0$.}
\begin{align}
J^n_{MN}(k)\equiv \frac{1}{V} \sum_{l }\int  \,d^4 p \,
\frac{\delta(pu')\delta(ku-pu)\, e^{-i(pb)}}{ (p^2-\ull^2)\,
[(k-p)^2 - (\uln^i - \ull^i)^2] }\, p_{M}  p_{N}\,,
\end{align}
are computed in Appendix \ref{app2}.  Using (\ref{cucu1_5}),
(\ref{susu0_5}) and (\ref{iij1}) one obtains\footnote{We passed
again  from summation into integration. Also, to simplify notation
a bit, the index ($n$), labeling the number of emission KK-mode,
is suppressed.}
\begin{align}
\label{gg4f} S_{MN}(k)= \frac{ \lambda \, e^{i(kb)}}{v} \int
\limits_0^1 &\left[ \frac{b_M b _N}{b^2}\gamma
\Gamma\hat{K}_{d/2+1} (\zeta_n)+\left(2i (ku') u_{(M} b_{N)}
+\frac{\Gamma}{\gamma v^2}\left[ 2i \gamma^2 v^2
b^{\vphantom{*}}_{(M}\tilde N_{N)} -  u_M u_N
 \right] \right)\hat{K}_{d/2} ( \zeta_n)+\right.\nn\\ & \left.
 +b^2\left(\Gamma\gamma \tilde N_M
\tilde N_N+\frac{1}{\gamma}(ku')^2 u_M u_N+2 (ku') u_{(M}\tilde N_{N)}\right)\hat{K}_{d/2-1}(\zeta_n)
\right] \e^{-i(kb)x} dx \,,
\end{align}
with $\zeta_n$ given in (\ref{cucu2}). Upon substitution of
$\tilde N_{M}\equiv  -\left[ (1-x)(ku) + \gamma x(ku')  \right] u_M/(\gamma^2 v^2)$,
one ends-up with the fully expanded expression
\begin{align}
\label{Sstress}
& S_{MN} \!= \! \frac{\lambda \,e^{i(kb)}}{v}
\!\! \int \limits_0^1 \! \left\{\! \frac{b_M b _N}{b^2}\gamma
\Gamma\hat{K}_{d/2+1} (\zeta_n) \!+ \! \! \left[  2 i
{u}_{(M}b_{N)}  \!  \left(  \! (ku') \! \left(\! 1-\frac{\tD
x}{v^2} \!\right) \! -\frac{\Gamma}{\gamma v^2} (1-x)(ku)\!
\right) \!-\! \frac{\Gamma}{\gamma v^2}  u_M u_N
 \! \right] \! \hat{K}_{d/2} ( \zeta_n)\right.\nn\\ & \left.
 \!+\!\left[ (ku')^2\! \left( \! 1\!-\!\frac{2x}{v^2}\!+\!\frac{\Gamma x^2}{v^4} \! \right)\!
 -2\frac{(ku)(ku')(1-x)}{\gamma v^2} \! \left(\! 1-\frac{\tD x}{v^2} \!\right) \!
 +\frac{\tD (ku)^2 (1-x)^2}{\gamma^2 v^4} \! \right] \! \frac{ b^2 }{\gamma} u_M u_N \hat{K}_{d/2-1}(\zeta_n)
\right\}  \e^{-i(kb)x} dx \,.
\end{align}
Like the local source (\ref{Tloc}), the non-local $S_{MN}$ in its effective form
(\ref{Sstress}) is also confined on the brane.
Furthermore, $S_{MN}$ is a linear combination of terms
proportional to integrals of the form
\begin{align}
\label{Jst}
{J}_{(\sigma, \tau)} \equiv \Lambda_d \int\limits_0^1 x^{\sigma}\, e^{-i(kb)x}
 \hat{K}_{d/2+\tau}(\zeta_n) \, dx\,,
\end{align}
(with $\sigma=0,1,2$), defined in (\ref{iii4}) and studied in
Appendix \ref{app2}\footnote{Note that the coefficient
$\Lambda_d$, which appeared naturally in the definition
(\ref{iii4}) of Appendix \ref{app2}, is kept here also, in order
to avoid the introduction of more symbols. Also, note that
$J_{(\sigma,\tau)}$ are clearly functions of $k$ and $n$, which
are suppressed in order to simplify notation a bit.}. In
\cite{GKST-3} it was shown that these integrals can be written as
a sum of two terms. One, is expressed in terms of hatted Macdonald
functions of only the variable $z$ and the other hatted Macdonalds
of another variable $z'$. So, one writes in an obvious notation
\begin{equation}
\label{J[z]} J_{(\sigma, \tau)} = J^{[z]}_{(\sigma, \tau)} +
J^{[z']}_{(\sigma, \tau)} \,, \qquad z\equiv \frac{(ku)b}{\gamma
v}\,, \qquad z'\equiv \frac{(ku')b}{\gamma v}\,.
\end{equation}
The total radiation amplitude is thus naturally written as a sum of two terms. Both expressed in terms of Macdonald
functions, but with argument $z$ the first and $z'$ the second.
As explained in \cite{GKST-3} the characteristics of the Macdonald functions translate eventually
into angular and frequency distribution properties of the emitted radiation.

\subsection{Destructive interference of the beamed high-frequency components}\label{di}

As explained in \cite{GKST-2} and \cite{GKST-3}, if the radiation
source consisted of the local part $T_{MN}$ alone, which is
expressed in terms of Macdonalds with argument $z$, the emitted
radiation would be mainly {\it z-type}, i.e. beamed $\vartheta
\sim \gamma^{-1}$, with high characteristic frequency $\od \sim
\gamma^2/b$. Now, it will be shown, exactly like in scalar
radiation discussed in \cite{GKST-3}, that in the sum
$T_{MN}+S_{MN}$ the two leading powers of $\gamma$ in the
ultra-relativistic expansions of $T_{MN}$ and $S_{MN}$ exactly
cancel and the total amplitude is suppressed, compared to each one
separately, by two powers of $\gamma$.  This cancelation is a
consequence of the equivalence principle: geodesics of the
ultrarelativistic particles and of the emitted massless quanta
stay close together and enhance the formation length of radiation
\cite{Khrip}.

We will not repeat here the derivation presented in \cite{GKST-3}. Instead, we will just highlight
the main steps of the derivation in the present context.

{\it Step 1}: For $z$-type radiation, one has $z\sim {\mathcal
O}(1)$ and $z'\sim {\mathcal O}(\gamma)$. Thus, the contribution
of $J^{[z']}$, which contains only Macdonalds of $z'$ are
exponentially suppressed and can be neglected. Thus, the dominant
contribution to $z$-type radiation is due to $J^{[z]}$.

{\it Step 2}: The integrals appearing in $S_{MN}$ with
$\sigma=1, 2$ are suppressed compared to the one with $\sigma=0$. Consider for instance
${J}_{(1,\tau)}$.  Using (\ref{J01pr}, \ref{reduct02}) and
(\ref{reduct2_3}) from Appendix \ref{app2} one can write for its
leading terms
\begin{align}
\label{DISA0} {J}_{(1,\tau)}^{[z]}=  \frac{\Lambda_{d}}{\coa^2}
\left(- \sin^2 \!\vartheta
\,\hat{K}_{d/2+\tau+1}(z)-\frac{(d+2\tau+3)\beta^2}{a^4 \coa^2}
\hat{K}_{d/2+\tau+2}(z) \! + \frac{\beta^2}{a^4
\coa^2}\hat{K}_{d/2+\tau+3}(z) \right) +\pp
\end{align}
with $a, \beta$  and $\xi$ defined in Appendices \ref{app1} and \ref{app2}. This is of
order of $a^{-2}\ll 1$ compared to $J_{(0,\tau)}^{[z]}$ (given in Appendix \ref{app2}), since in the
region of interest $a=\mathcal{O}(\gamma)$.

{\it Step 3}: Since the destructive interference effect has to do with the two leading terms in the
local and non-local amplitudes, we focus on exactly these two.
Hence, we neglect $(ku)/\gamma$
compared to $(ku')$, set $\Gamma =1$ and $v=1$ and omit $x-$ and $x^2-$terms in the integrand
of (\ref{Sstress}).
Thus, the leading part of $S_{MN}$ reads
\begin{align}
\label{S_in_z} S_{MN} \!\simeq \! \lambda e^{i(kb)}\!\! \int
\limits_0^1 \! & \left[\! \frac{b_M b _N}{b^2}\gamma
 \hat{K}_{d/2+1} (\zeta_n) \!+  \! \left(  2
i  {u}_{(M}b_{N)}   (ku')  \! -\! \frac{1}{\gamma}  u_M u_N
 \! \right) \! \hat{K}_{d/2} ( \zeta_n)
 \!+ \!  {z'}^2 \! \gamma \hat{K}_{d/2-1}(\zeta_n) u_M u_N \!
\right]\! \e^{-i(kb)x} dx\,.
\end{align}

{\it Step 4}: The quantities to be compared are the projections of
$S_{MN}$ and $T_{MN}$ on the polarization tensors. It is easy to
check, using the formulae of Appendix \ref{app1} for the
contractions of $u^M$ and $b^M$ with the polarization vectors,
that at small angles $\vartheta \lesssim \gamma^{-1}$ no extra
powers of $\gamma$ are introduced through such contractions. Thus,
comparing directly the coefficients of $u_M u_N$, ${u}_{(M}b_{N)}$
and $b_M b _N$, one ends-up with the following expression for the
leading terms of $S_{MN}$:
\begin{align}
\label{S_in_z1} S_{MN} \simeq \lambda e^{i(kb)}\!\!
\int\limits_0^1 \! & \left[   2i {u}_{(M}b_{N)}   (ku')
\hat{K}_{d/2} ( \zeta_n)
 +  {z'}^2  \gamma \hat{K}_{d/2-1}(\zeta_n) u_M u_N
\right] \e^{-i(kb)x} dx  ,
\end{align}

{\it Step 5}: Using (\ref{J0}) appropriately simplified for the range of parameters of interest here, namely
$$\int
\limits_0^1  dx \, \e^{-i(kb)x}\, \hat{K}_{\tau-1}(\zeta_n) \simeq \frac{1}{\gamma z z'} \hat{K}_{\tau}(z)-i
\frac{ (kb)}{(\gamma z z')^2}\hat{K}_{\tau+1}(z)+ \pp \,,
$$
into (\ref{S_in_z1}), one obtains
\begin{align}
\label{S_in_z3}
S_{MN} \simeq  \lambda e^{i(kb)}
 \left[  \frac{z'}{z} \hat{K}_{d/2}(z) u_M u_N  -i
\frac{1}{\gamma z^2}\hat{K}_{d/2+1}(z)   \sigma_{MN}
\right]\,,
\end{align}
plus contributions to the real and the imaginary parts suppressed by two powers of $\gamma$
compared to the terms kept in (\ref{S_in_z3}).

{\it Step 6}: Making the same approximations for the local part $\un T_{MN}$ in (\ref{Tloc}), one ends-up
with $^1T_{MN}=-S_{MN}$ given in (\ref{S_in_z3}) above and proves the so called {\it destructive interference}
of the leading terms of local and stress amplitudes.

\subsection{The total radiation amplitude}\label{tra}

To summarize, the total radiation source $\tau_{MN}$ was naturally split into the sum of a function of $z$ and
a function of $z'$. The two leading terms in the expansion of the function of $z$ in powers of $\gamma$
were shown to cancel.
The next step is to compute the leading terms of $\tau_{MN}$, which survive destructive interference and use them to
compute the emitted radiation energy to leading order. The discussion follows the steps described in
\cite{GKST-3} for scalar radiation, only repeated three times because of the three relevant polarizations. That is, for
each polarization one first distinguishes three characteristic regimes of angular
and frequency distribution of the emitted energy, namely (a) the beamed ($\vartheta\sim 1/\gamma$) high-frequency
($\omega\sim \gamma^2/b$) regime, (b) the beamed medium-frequency ($\omega\sim \gamma/b$) regime, and
(c) the unbeamed $\vartheta\sim 1$ medium-frequency regime, and then computes the emitted energy in
each one of them.

\subsubsection{The beamed, high-frequency radiation amplitude}

In this regime $z'\sim {\mathcal O}(\gamma)$ and the contribution of Macdonald functions of $z'$ is exponentially
suppressed.

\vspace{0.2cm}

\textbf{Polarization I:} Projecting (\ref{Tloc}) and (\ref{Sstress}) on $\fip$ and taking into account that only the
$e_1^{M}e_1^{N}/\sqrt{2}$ part of $\fip$  contributes, one obtains for $^1T_{\rm I}\equiv \! {\un}T_{MN}\fip$ and
$S_{\rm I}\equiv S_{MN}\fip$
\begin{align}
\label{TlocI}
\un T_{\rm I}(k)=- \frac{ \lambda \, e^{i(kb)}}{\sqrt{2}\gamma
v^3} \left[-i \Gamma \frac{\hat{K}_{d/2+1}(z)}{z^2} (kb) [\psi^2
\gamma^2 -1]+ \!\left(\! \left(2v^2 \!- \! \Gamma\right)\gamma
\frac{z'}{z} -\Gamma\! \right) \hat{K}_{d/2}(z) \gamma^2 v^2
\sin^2 \vartheta\right]\,,
\end{align}
and
\begin{align}
\label{SstressI} & S_{\rm I} \!= \! \frac{\lambda \gamma
\,e^{i(kb)}}{\sqrt{2 } v} \!\! \int \limits_0^1 \! \left\{\!
(\cos^2 \phi \cos^2\vartheta- \sin^2 \phi) \Gamma\hat{K}_{d/2+1}
(\zeta_n) \!+ \! \! \left[  2i   v    \cos \vartheta (kb)
 \left(\! 1-\frac{\tD x}{v^2}
-\psi \frac{\Gamma(1-x)}{  v^2} \! \right) \!-\!  \Gamma
   \sin^2 \vartheta
  \right] \times\right.\nn\\ & \left. \times  \hat{K}_{d/2} ( \zeta_n)
 \!+\!\left[  \frac{1}{\psi^2}\! \left( \! 1\!-\!\frac{2x}{v^2}\!+\!\frac{\Gamma x^2}{v^4} \! \right)\!
 -2\frac{ (1-x)}{ \psi v^2} \! \left(\! 1-\frac{\tD x}{v^2} \!\right) \!
 +\frac{\tD   (1-x)^2}{ v^4} \! \right] \! z^2  v^4 \sin^2 \vartheta \hat{K}_{d/2-1}(\zeta_n)
\right\}  \e^{-i(kb)x} dx \,,
\end{align}
respectively. The next step is to expand in powers of $\gamma$. For that one has to use the expansions of
$\cos\vartheta$, $v$, $\Gamma$ and $\psi$ in Appendix \ref{app1}, the fact that integrals with extra $x^\alpha$ in
the integrand are suppressed by $\gamma^{-2\alpha}$, and the expressions for the integrals $J^{[z]}_{(0,\tau)}$ and
$J^{[z]}_{(1,\tau)}$ given in Appendix \ref{app2} and (\ref{DISA0}). After a somewhat tedious but straightforward
calculation one ends-up with the leading part of $\tau_{\rm I}\equiv \tau_{MN}\fip$:
\begin{align}
\label{tauI} &  \tau_{\rm I} \! \simeq   \frac{\lambda
\,e^{i(kb)}}{\sqrt{2 } } \frac{ \sin^2
 \vartheta }{\gamma  \psi}  \left[\frac{d+1}{d+2}
\hat{K}_{d/2}(z)
 +    (d+1)
\frac{\hat{K}_{d/2+1}(z)}{z^2}+ \left(\frac{\gamma^2 \psi^2 \cos 2
\phi}{ \sin^2
 \vartheta } - \sin^2\phi
  \right)\frac{\hat{K}_{d/2+2}(z)}{z^2}
\right] \,.
\end{align}

\textbf{Polarization II:} Similarly, projection onto the second polarization tensor leads (with an analogous notation)
to
\begin{align}
\label{TlocII} \un T_{\rm II} =  \frac{ 2  i  \lambda \,
e^{i(kb)}}{\sqrt{2}  v }   \frac{\hat{K}_{d/2+1}(z)}{z } \gamma
\sin \vartheta \sin\phi \,,
\end{align}
and
\begin{align} S_{\rm II}  \!= \! \frac{ \gamma \lambda \,e^{i(kb)}}{\sqrt{2}v}
\!\! \int \limits_0^1 \! \left[\Gamma \cos \vartheta \sin 2 \phi
\hat{K}_{d/2+1} (\zeta_n)  - 2 i  z  \sin\vartheta \sin\phi \left(
\frac{v^2}{\psi}\left(\! 1-\frac{\tD x}{v^2} \!\right) \! - \Gamma
(1-x)  \right)
  \hat{K}_{d/2} ( \zeta_n)
\right] \e^{-i(kb)x} dx \,. \nn
\end{align}

Following the same steps as above, one obtains for the leading ultra-relativistic term in their sum
\begin{align}
\label{tauII}  \tau_{\rm II}  \simeq  \frac{ \lambda
\,e^{i(kb)}}{\sqrt{2} }  \gamma \left(  \psi  + \sin^2\vartheta
\right) \sin 2 \phi   \frac{\hat{K}_{d/2+2} (z)}{z^2}   \,.
\end{align}

\textbf{Polarization III:} Repeating the same steps for the third
polarization tensor, while taking into account that only the $b_M
b_N-$binomial does not vanish upon projection on $e_2^M e_2^N$,
one obtains:
\begin{align}
\label{TlocIII} \un T_{\rm III} = -\sqrt{\frac{d}{d+2}} \un T_{\rm
I}\, ,
\end{align}
and
\begin{align}
\label{SstressIII}S_{\rm III} = -\sqrt{\frac{d}{d+2}} \left[S_{\rm
I}+ \frac{\lambda \gamma \,e^{i(kb)}}{\sqrt{2}} \!\! \int
\limits_0^1 2\sin^2 \phi \hat{K}_{d/2+1} (\zeta_n)\e^{-i(kb)x}
dx\right]\, .
\end{align}
Combining these two and expanding, one has to leading order
\begin{align}
\label{tauIII} &  \tau_{\rm III} \!= -\sqrt{\frac{d}{d+2}}
  \frac{\lambda \,e^{i(kb)}}{\sqrt{2 } } \frac{
\sin^2
 \vartheta }{\gamma  \psi}  \left[\frac{d+1}{d+2}
\hat{K}_{d/2}(z)
 +    (d+1)
\frac{\hat{K}_{d/2+1}(z)}{z^2}+ \left(\frac{\gamma^2 \psi^2  }{
\sin^2
 \vartheta } - \sin^2\phi
  \right)\frac{\hat{K}_{d/2+2}(z)}{z^2}
\right] \,.
\end{align}
Up to the common phase factor and the factor $\lambda$ in front, all three amplitudes are real and
of ${\mathcal O}(1/\gamma)$.

\subsubsection{Wide-angle, medium-frequency radiation amplitude}

In this regime it is more convenient to consider directly the behavior of the radiation source, before one projects
onto the three polarization tensors. Also, in this regime $z\sim \gamma$, and consequently the Macdonald functions of
$z$ give exponentially suppressed contributions and will be neglected.
Thus, the leading contribution is due to the $z'-$dependent part of the non-local source $S_{MN}$, which we compute
next.

In Subsection \ref{di} it was shown that for $\od \sim \ga^2/b, \; \vartheta \sim \gamma^{-1}$ one has
$J_{(1,\tau)} \sim J_{(0,\tau)}/\gamma^2.$ This implied that only the neighborhood of $x=0$ contributes
to these integrals.

Similarly, it will be shown that in the region of interest here, namely ($\od \sim \ga/b, \; \vartheta \sim 1$), the leading
contribution to these integrals comes from the neighborhood of $x=1$.
Indeed, consider the integral with $(1-x)$, i.e. $J_{(0,\tau)} -J_{(1,\tau)} $. Using (\ref{reduct02}) and subsequently
(\ref{J01pr}) this difference takes the form
\begin{align}\label{re01b}
{J}_{(0,\tau)}-{J}_{(1,\tau)} \simeq \frac{\Lambda_d\, \e^{-i (k
b)}}{\xi^2} \left[ \left(1-\frac{2}{\psi}
\right)\frac{\hat{K}_{d/2+\tau+1}(z')}{\gamma^2} + \frac{\cos^2 \!
\phi}{a^2} \hat{K}_{d/2+\tau+3}(z')+\pp\right].
\end{align}
This is of order $\mathcal{O}(\ga^{-2})$ with respect to both
${J}_{(0,\tau)}$  and  ${J}_{(1,\tau)}$. This implies that the
integrals ${J}_{(0,\tau)}$ and  ${J}_{(1,\tau)}$ are almost equal,
and the main contribution in them comes from the domain $x=1-0$,
and consequently, that integral with $(1-x)^2$ is even more
suppressed.

Thus, it is natural to rearrange the terms of the integrand in (\ref{Sstress}) in powers of $1-x$ as well as in
powers of $\gamma$, taking into account that each power of $1-x$ contributes a factor of ${\mathcal O}(1/\gamma^2)$
to the integral. Taking in addition into account that in the region of
interest here $\gamma(ku')$ is of the same order as $(ku)$, one obtains
\begin{align}\label{Sstress2}
\tau_{MN} \!\simeq     \lambda \, e^{i(kb)}  \!\! \int \limits_0^1
\! & \left\{\! \vphantom{\frac{{z'}^2}{\gamma}} \frac{b_M b
_N}{b^2}\gamma\hat{K}_{d/2+1} (\zeta_n) \!+ \! \! \left[  2 i
{u}_{(M}b_{N)}  \!  \left(  \!
\left(\!(ku')\!-\!\frac{(ku)}{\gamma}\right) \left( 1-x\right) \!
-\frac{d+1}{d+2}\frac{(ku')}{\gamma^2}\! \right) \!-\! \frac{u_M
u_N}{\gamma }
 \! \right] \! \hat{K}_{d/2} ( \zeta_n)\right.\nn\\ &\;\; \left.
- \frac{1}{d+2} \frac{{z'}^2}{\gamma} u_M u_N \hat{K}_{d/2-1}(\zeta_n)
\right\}  \e^{-i(kb)x} dx \,.
\end{align}

Using $(e_1u)=\gamma v \sin\vartheta$, $(e_2u)=0$ and $(e_1b)={\mathcal O}(b)=(e_2b)$,
one can simplify (\ref{Sstress2}) and write instead
\begin{align}
\label{Sstress3}
\tau_{MN} \!\simeq   \frac{ \lambda \, e^{i(kb)} }{\gamma}
\int \limits_0^1 \! \left[ \frac{b_M b _N}{b^2}\gamma^2
\hat{K}_{d/2+1} (\zeta_n)  -  u_M u_N \hat{K}_{d/2} ( \zeta_n) -
\frac{1}{d+2}  {z'}^2  u_M u_N \hat{K}_{d/2-1}(\zeta_n) \right]
\e^{-i(kb)x} dx \,.
\end{align}
It is straightforward to check that its projection onto any of the three polarization tensors gives to leading order
the same result as the projection of (\ref{Sstress2}).

From (\ref{J01pr}) one obtains to leading ultra-relativistic order
\begin{align}
\label{J01pr_largeangle}
\int\limits_0^1 \e^{-i(kb)x}  \hat{K}_{\tau} ( \zeta_n)\, dx \simeq \frac{\e^{-i (k b)} }{{z'}^2 \ga^2 \psi}
\hat{K}_{\tau+1}(z') \, ,
\end{align}
using which in (\ref{Sstress3}), one finally ends-up with
\begin{align}
\label{Sstress4}
\tau_{MN}  \simeq   \frac{\lambda}{{z'}^2 \gamma^3 \psi}
\left[ \frac{b_M b _N}{b^2}\gamma^2 \hat{K}_{d/2+2} (z')  -  u_M
u_N \hat{K}_{d/2+1} (z') - \frac{1}{d+2}  {z'}^2  u_M u_N
\hat{K}_{d/2}(z') \right] .
\end{align}

\subsubsection{The beamed, medium-frequency radiation amplitude}

Starting again from (\ref{Tloc}) and (\ref{Sstress}), one can
observe that for $\vartheta \sim \gamma^{-1}$ the projection on
polarization tensors does not introduce $\gamma$-factors, hence
one can estimate the order in $\gamma$ directly from the
coefficients of the tensor binomials $u_M u_N$, $b_M b_N$ and
$u_{(M}b_{N)}$. In the regime of interest here one has $z \sim
1/\gamma, \, z' \sim 1$, using which one can immediately conclude
that both local and stress sources, as well as their projections
on the relevant polarizations are of the same order, namely
$T_{MN} \sim S_{MN} \sim \tau \sim \mathcal{O}(\gamma)$ in any
dimension.

\subsection{Summary}\label{summ}
The results of this section are summarized in Table I below, which shows the
leading behavior of the corresponding amplitudes after projection on the polarization tensors.

\vspace{0.3cm}

\noindent
\begin{tabular}{|c|c|c|c|}\hline
  \backslashbox{$\,\vartheta\!$}{$\omega$}  &  $\omega \sim 1/b $ & $
  \omega \sim \gamma/b $& $\omega \sim
  \gamma^2/b$ \\ \hline
  $\gamma^{-1}$ &  $\begin{array}{l}\text{\small  no destructive interference} \\
   \tau \sim T \gg S  \end{array}$   &
    $ \begin{array}{l}\text{\small  no destructive interference} \\ S^{[z]} \sim T  \sim S^{[z']}
   \sim \gamma  \end{array}$   &
  \ths $\begin{array}{l} \text{\small  destructive interference: }T \approx -S^{[z]} \\S^{[z']}\sim \exp(-\gamma), \\
   \tau =\mathcal{O} (T/\gamma^2) \sim 1/\gamma \end{array} $
  \ths  \\[10pt] \hline
  1 &
  \ths$\begin{array}{l}\text{\small  no destructive interference} \\ \tau \sim T \sim S
  \end{array}$ \ths &
   \ths $\begin{array}{l} \text{\small  destructive interference: } \\  S^{[z]} \approx T
   \sim \exp(-\gamma)\\ \tau=S=S^{[z']}\sim \gamma^{-1}  \end{array}$ \ths  &
   \ths $\begin{array}{l} \text{\small  destructive interference  } \\ T \sim S \sim \tau
   \sim \exp(-\gamma)
    \end{array} \ths$
\\[10pt] \hline
\end{tabular}\\

{\small Table I. The leading behavior of the amplitudes (projected on polarizations)
in the various characteristic angular-frequency regimes. $S^{[z]}$ and
$S^{[z']}$ stand for the contributions of integrals $J^{[z]}$ and $J^{[z']}$, respectively.
The dependence of $\lambda$ on $\gamma$ is not included, i.e.
all estimates are given in units $\lambda=1$.}

\vspace{0.5cm} Notice that the entries in Table I do not depend on
the dimensionality $d$. Also, note that this leading dependence of
the amplitudes on $\gamma$ is valid in the whole range $[0, 2\pi]$
of the azimuthal angle $\varphi$.

\section{The emitted energy}

The emitted energy, obtained from (\ref{DEMD}) for $M=0$ is
\begin{align} \label{DEMD_e} E= \frac{\varkappa_D^2}{4(2
\pi)^{D-1}} \sum_{\cp } \int\limits_{0}^{\infty}\od^{D-2} d\od
\!\! \int\limits_{S^{D-2}} \!\! d\Omega \; \left|
\tau_{LR}(k)\,\varepsilon^{LR}_{\cp}\right|^2 .
\end{align}
It differs from the scalar radiation case studied in \cite{GKST-3} by the summation over polarizations and the
substitution of the scalar charge $f \to \vk m$. Thus, it is expected to lead up to numerical coefficients of
order $\mathcal{O}(1)$ to the same behavior in $\gamma$ as the energy emitted in the scalar case.
On the basis of the qualitative arguments given in the beginning of Section 4 of \cite{GKST-3}
and the subsequent numerical computations one obtains\footnote{Up to a logarithmic overall factor for
$d=1$, discussed below.} for the emitted energy in the various
angular-frequency domains the behavior summarized in Table II.

\vspace{0.3cm}
\begin{tabular}{|c|c|c|c|c|}
\hline \backslashbox{$\vartheta$}{$\omega_D$} & $\omega_D\ll \gamma/b$  &
$ \omega_D \sim \gamma/b$ & $\omega_D \sim \gamma^2/b $& $\omega_D \gg \gamma^2/b$
  \\\hline
  $\gamma^{-1}$ & $\begin{array}{c} \text{\small negligible}\\ \text{\small  (phase space)} \end{array}$ &
  $\begin{array}{l} \hspace{0.1cm} E \sim \gamma^3\,, \text{\;\;from $T$ and $S$ \hspace{0.0cm}}
  \\ \end{array}$  &
   $\begin{array}{l} \hspace{0.3cm} E \sim \gamma^{d+2}, \text{\; from $T+S^{[z]}$ \hspace{0.2cm} }\\
   \end{array}$
   & $\begin{array}{c}  \text{\small  negligible radiation}\\ \end{array}$
  \\[20pt]   \hline
  1 & $\begin{array}{c} \text{\small negligible}\\ \text{\small  (phase space)} \end{array} $ &
   $\begin{array}{l} E \sim \gamma^{d+1}, \text{\;\;from $S^{[z']}$ }\\   \end{array} $&
   $\begin{array}{c}  \text{\small negligible radiation}\\  \end{array}$&
    $\begin{array}{c}  \text{\small  negligible radiation}\\  \end{array} $\\[20pt] \hline
\end{tabular}\\

\vspace{0.0cm}
{\small Table II. The relative contribution in the total emitted energy
from different characteristic regions of angle and frequency. All
estimates are given in units $\lambda=b=1$.} \label{table2}

\vspace{0.3cm}


\subsection{The total radiated energy}

According to Table II the frequency and angular distribution of
the dominant component of radiation depends on the dimensionality
$d$. Thus, like in the scalar case \cite{GKST-3} the study of the
emitted energy has to proceed separately for $d\geqslant 2$ and
for $d=0, 1$. Furthermore, the following analysis is a triple
repetition of the scalar case, because of the three polarizations.

\vspace{0.3cm}

\textbf{$\boldsymbol{d \geqslant 2}.$}

For $d\geqslant 2$ the dominant radiation is beamed with
characteristic frequency around $\omega \sim \ga^2/b$. According
to Table I, the emission amplitude is $\mathcal{O}(\gamma^{-1})$,
dominated by its real part.

For $R\gg b$ the KK-summation is replaced by integration and the relevant formula for the emitted
energy is (\ref{DEMD}). One next substitutes the high-frequency amplitudes (\ref{tauI},
\ref{tauII} and \ref{tauIII}) corresponding to the three polarizations and neglects their
behavior at smaller frequencies. Squaring each amplitude, one recognizes the appearance of products
of $K_{d/2}(z)$, $\hat{K}_{d/2+1}/z^2$ and $\hat{K}_{d/2+2}/z^2$ with themselves, and also the presence
of the factor $\omega^{d+2}$ from phase space.
Thus, one starts with
\begin{align} \label{DEMD_ef} \frac{dE}{d\Omega}= \frac{\varkappa_D^2}{4(2
\pi)^{D-1}} \sum_{\cp } \int\limits_{0}^{\infty}\od^{D-2} d\od
 \left| \tau_{LR}(k)\,\varepsilon^{LR}_{\cp}\right|^2  ,
\end{align}
and upon integration over the angles except $\vartheta$ one obtains
\begin{align}
\label{CabDab_theta_z}
\frac{dE}{d \vartheta} = \frac{ (\varkappa_D^3 m m')^2
\gamma^{d+1}}{8 \, (2 \pi)^{2d+5}   \, b^{3d+3} } \frac{\sin^{d+3}
\vartheta}{ \psi^{d+3}} \sum_{a,b=0}^2 C_{ab}^{(d)}
D_{ab}^{(d)}(\vartheta)\,.
\end{align}
where
\begin{equation}\label{Cab}
C_{ab}^{(d)} \equiv \int\limits \hat{K}_{d/2+a}(z)\hat{K}_{d/2+b}(z)
z^{d+2(\delta_{0a}+\delta_{0b}-1)} dz\,,
\end{equation}
\begin{align}\label{Dab_theta_z}
& D_{00}^{(d)}(\vartheta)= 2\frac{(d+1)^3}{(d+2)^3}\frac{\sin^2\!\vartheta}{\gamma^2 \psi^2} &
& D_{01}^{(d)}(\vartheta)=(d+2)  D_{00}^{(d)}(\vartheta)  \nn \\
& D_{02}^{(d)}(\vartheta)= \frac{d+1}{(d+2)^2} \left[d  -(d+1)\frac{\sin^2\!\vartheta}{\gamma^2 \psi^2}\right] &
 & D_{11}^{(d)}(\vartheta)= (d+2)^2  D_{00}^{(d)}(\vartheta) \nn \\
& D_{12}^{(d)}(\vartheta)=  (d+2)D_{02}^{(d)}(\vartheta) & &
D_{22}^{(d)}(\vartheta)=\frac{d+1}{d+2}\left( 2 \frac{\gamma^2
\psi^2}{\sin^2\!\vartheta}+\frac{3}{4}
\frac{\sin^2\!\vartheta}{\gamma^2 \psi^2}
 \right)  +\frac{2}{d+2}+\gamma^2\sin^2\!\vartheta \, .
\end{align}
(all of ${\mathcal O}(1)$ for $\vartheta\sim 1/\gamma$), while use was made of (\ref{intSN}):
\begin{equation} \int \limits_{S^{d+1}} \!\! d\Omega_{d+1} =\Omega_{d+1}\;, \quad
\int \limits_{S^{d+1}} \!\! \cos^2 \! \phi \, d\Omega_{d+1}
=\frac{1}{2}\Omega_{d+1}\; , \quad \int \limits_{S^{d+1}}\!\! \cos^4
\! \phi \, d\Omega_{d+1} =\frac{3}{8}\Omega_{d+1}\; .
\label{angularintegrals}
\end{equation}
Upon integration over $\vartheta$ using (\ref{jj2q}) one ends up with
\begin{align}\label{CabDab_z}
E = \frac{\pi^{d/2+1} (\varkappa_D^3 m m')^2  \gamma^{d+2}}{8 \,
(2 \pi)^{2d+5} \Gamma(d/2+1) \, b^{3d+3} }
    \sum_{a,b=0}^2 C_{ab}^{(d)} D_{ab}^{(d)}\, ,
\end{align}
where the constants $D_{ab}^{(d)}$ are given by
\begin{align}\label{Dab_z}
 & D_{00}^{(d)} = 2^{d+5}\frac{(d+1)^3}{(d+2)^3}\frac{\Gamma\left(\frac{d+6}{2}\right)
 \Gamma\left(\frac{d+4}{2}\right)}{\Gamma(d+5)}  & & D_{01}^{(d)} =(d+2)  D_{00}^{(d)}  \nn \\
  & D_{02}^{(d)} =-2^{d+3}\frac{(d+1)(2d+3)}{(d+2)^2} \frac{\Gamma\left(\frac{d+4}{2}\right)
  \Gamma\left(\frac{d+2}{2}\right)}{\Gamma(d+4)}   & & D_{11}^{(d)} = (d+2)^2  D_{00}^{(d)}  \\
 & D_{12}^{(d)} = (d+2)D_{02}^{(d)}  & & D_{22}^{(d)}
 =2^{d-2}(5d+6)(3d^2+15d+20)\frac{\Gamma\left(\frac{d+2}{2}\right)\Gamma\left(\frac{d}{2}\right)}{\Gamma(d+4)} \, . \nn
\end{align}
 Finally, after the summation in (\ref{CabDab_z}) one obtains for the total emitted energy in the lab frame
\begin{align}\label{Energy_emit_z}
 E \simeq C_d \frac{(\varkappa_D^3 m m')^2}{b^{3d+3} } \gamma^{d+2}
\end{align}
with $C_2=6.23\times 10^{-4}$, $C_3=2.61\times 10^{-4}$,
$C_4=1.98\times 10^{-4}$, $C_5=2.16\times 10^{-4}$,
$C_6=3.10\times 10^{-4}$.


\vspace{0.3cm}

\textbf{$\boldsymbol{d=1}.$}

The emitted energy flux as well as its frequency and angular distributions in $d=0, 1$ were
computed numerically.

According to Tables I and II in this
case most of the radiation is beamed $\vartheta\lesssim 1/\gamma$,
but, as it was described in \cite{GKST-3} and immediately follows
from the destructive interference, in five dimensions the
frequency distribution falls as $1/\omega$ in the region from
$\mathcal{O}(\ga/b)$ to $\mathcal{O}(\ga^2/b)$, hence the total
emitted energy grows as $\gamma^3 \ln \gamma$, while the dominant
contribution comes from this entire region. From the known expressions for the
amplitudes in integral form, we compute the integral numerically for several
values of $\gamma$ and conclude:
\begin{align}\label{Energy_emit_z_5D}
 E \simeq C_1 \frac{(\varkappa_5^3 m m')^2}{b^{6} } \gamma^{3}
 \ln\gamma \;, \qquad C_1 =1.66 \times 10^{-4}.
\end{align}

\vspace{0.3cm}

\textbf{$\boldsymbol{d=0}.$}

\begin{figure}
\centering
 \subfigure[ ]{\raisebox{0pt}{
\includegraphics[ width=8.2cm]{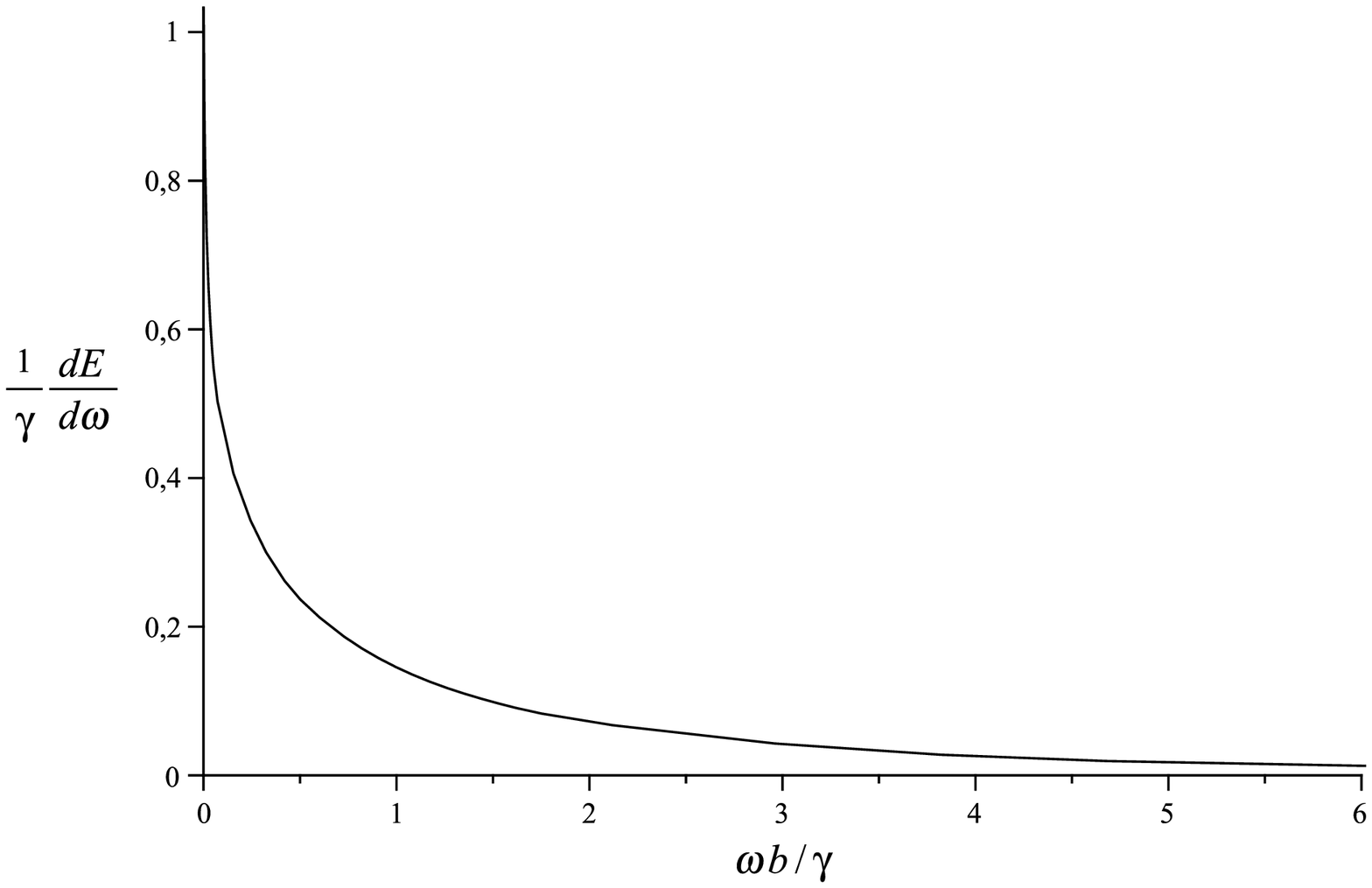}\label{4D_fr_dist}}}
\subfigure[]{ \raisebox{1pt}{
\includegraphics[width=8.2cm]{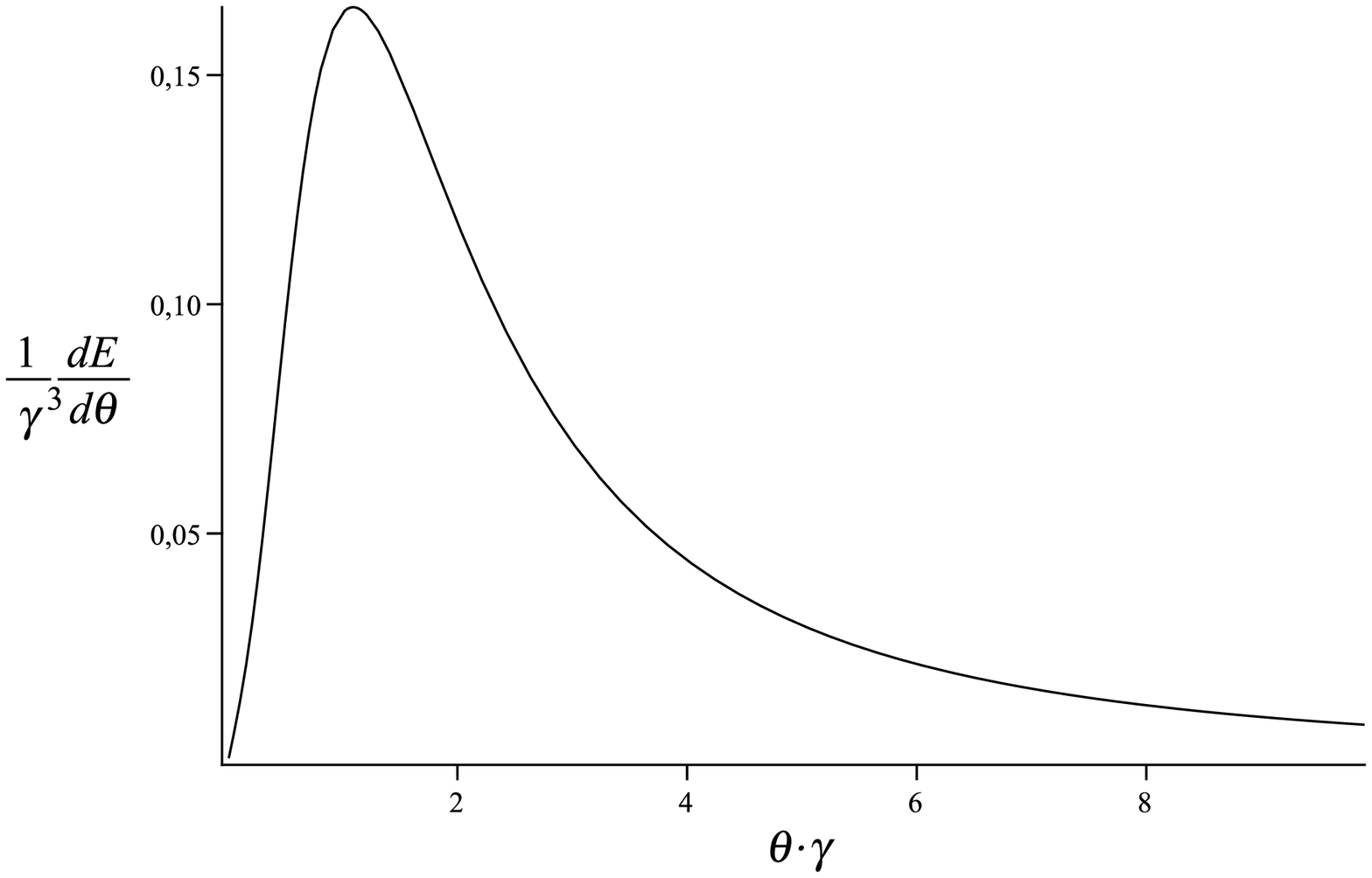}\label{4D_ang_dist}}}
\caption{Frequency (a) and angular (b) distribution of the total
emitted energy for $d=0$ and $\gamma=10^3$.} \label{4D_spectrum}
\end{figure}

Again, according to Tables I and II in this case most of
the radiation is beamed within $\vartheta\lesssim 1/\gamma$ and with
frequencies around $\omega \sim \gamma/b$, with the contribution from higher
frequencies decaying as $1/\omega^2$. In this domain the
local and non-local parts of the amplitude are equally important.
Thus, one has to add (\ref{Tloc}) and (\ref{Sstress})
to compute the total radiation amplitude, project onto the two
polarizations $\epsilon_{\pm}$, square each one of those, add them up
and multiply by the appropriate phase space factor. After the
angular integration one obtains the frequency (or $z'$)
distribution of the emitted energy shown in Figure
\ref{4D_spectrum}a,
while the angular distribution of the total emitted
energy (which carries the anisotropy) is given in Figure \ref{4D_spectrum}b.

Integrating the frequency distribution numerically over $\omega$
from 0 to $+\infty$ one obtains for the total emitted energy
\begin{align}
E \simeq C_0 \frac{( \varkappa_4^3 m m')^2}{b^3}\gamma^3 \;, \qquad
C_0=5.7\times 10^{-4}. \label{Edequal0}
\end{align}

\subsection{The low frequency part of the spectrum}
A few comments, related to the low frequency part of the classical radiation spectrum, are in order here.

(a) For $\boldsymbol{d=0}$ the distribution $dE/d\omega$ of the emitted energy has non-vanishing
finite zero-frequency limit.

Indeed, comparing local (\ref{Tloc}) and stress (\ref{Sstress})
sources one concludes that for $\omega \to +0$ the dominant part
of the amplitude is given by the imaginary part of the local piece
\footnote{In the diagrammatic quantum mechanical treatment of soft
gravitational radiation there is a corresponding statement, namely
that the soft radiation emitted from internal lines is negligible.
Only gravitons attached on the external lines contribute in the
zero-frequency limit.} (omit the phase factor $e^{i(kb)}$), whose
behavior is
\begin{align}\label{Tloc_lowfr}
 \tau_{MN}(k)\simeq -  i\frac{ \lambda \, \Gamma}{\gamma v^3}
 \frac{\hat{K}_{1}(z)}{z^2}\sigma_{MN} \sim \frac{1}{\omega}\, .
\end{align}
Upon contraction with the polarizations $\varepsilon_{\pm}$, substitution of the finite
zero-limit $\hat{K}_{1}(0)=1$ of the hatted
Macdonald function, and setting
$v=1, \; \tD=1$, one obtains
\begin{align}\label{Tloc_lowfr1}
 \tau_{\pm}\simeq  \frac{ i\lambda }{2}
 \frac{ \gamma \sin \vartheta}{\omega b  }\left[\cos \phi \left(\frac{1}{\gamma^2 \psi^2}-1\right)\pm 2i  \frac{ \sin
 \phi}{\psi} \right]\,.
\end{align}
Substitute this into (\ref{DEMD_ef}) and integrate over the
azimuthal angle $\phi$, to obtain
\begin{align}\label{Tloc_lowfr2}
\left.\frac{dE}{d\omega}\right|_{\omega=0} = \frac{\varkappa_4^2\lambda^2
\gamma^2}{ 16 (2\pi)^{2}   b^2}
 \int d \vartheta \,\sin^3 \!\vartheta \left(1 + \frac{4}{ \psi^2}+\frac{1}{ \ga^4 \psi^4} \right)
\end{align}
plus subleading powers of $\gamma$. Note that in the low frequency limit \textit{all} angles
($0\leqslant \vartheta \leqslant \pi$) contribute to $dE/d\omega$.

Finally, using $V_0^3\simeq  4/3$ (\ref{jj3}), $V_4^3\simeq
4\gamma^4/3 $ (\ref{jj2q}), $V_2^3 \simeq 4\,(\ln 2\gamma-1)$
(\ref{jj4}), and substituting the expression for $\lambda$, one ends-up with
\begin{align}\label{ZFL_4D}
\left.\frac{dE}{d\omega}\right|_{\omega=0} = \frac{({\varkappa}_4^3 m
m')^2 }{4(2\pi)^{4} b^2}\gamma^2 \left(\ln 2\ga
-\frac{5}{6}\right)\;,
\end{align}
which for $\varkappa_4=m=m'=b=1$ and $\gamma=10^3$ agrees with the
value obtained numerically in Figure \ref{4D_spectrum}a
\footnote{Note that this result differs from the one obtained in
\cite{Galtsov:1980ap}, where it was claimed that $dE/d\omega$
blows-up logarithmically as $\omega\to 0$.}.

(b) One can multiply the above result for $dE/d\omega|_{\omega=0}$ by the range
$\omega_{\rm max}\sim \gamma/b$ to obtain a rough estimate of the total emitted
energy. Its $\gamma$ dependence differs from (\ref{Edequal0}) by a logarithmic factor.

However, this simple method to obtain an order of magnitude estimate cannot be used in
$d>0$, because according to Table II the dominant
contribution to the emitted energy comes from the high frequency regime, which does not
have overlap with the $\omega\to 0$ domain.

(c) For $\boldsymbol{d>0}$, one cannot replace at small frequencies the KK-mode summation
with integration. One has instead to keep separately the three-dimensional coordinates
$\varpi,\theta,\varphi$ and the extra-dimensional discrete momenta $k_T$. In this case
$$z=\frac{(ku)b}{\gamma v}=\frac{b}{v}\left(\sqrt{ \varpi^2+k_T^2}- \varpi v \cos \theta \right)  \quad {\rm and} \quad
 z'=\frac{(ku')b}{\gamma v}=\frac{b}{\gamma v}\sqrt{ \varpi^2+k_T^2}\;,$$
while $(kb)= -\varpi \sin\theta \cos\varphi.$

The real part inside the square bracket of (\ref{gg2_5}) is regular or
blows-up logarithmically at $\varpi=0$. Thus, it vanishes when multiplied by
the volume measure.
Consequently, the dominant contribution to the emitted energy is due to the imaginary part, i.e. (ignoring the irrelevant
phase factor)
\begin{align}
\tau_{MN}(k) \simeq \un T_{MN}(k) \simeq \frac{i m m'
\varkappa_D^2}{4\pi V}\frac{1}{\gamma v^3} \!\sum_{l \in
\mathbb{Z}^d}  \frac{\hat{K}_{1}(z_l)}{z^2}\sigma_{MN} \,.
\end{align}
Taking into account that
$\hat{K}_{1}(z_l)$ is regular at all frequencies, the
low-frequency behavior is determined by the factor
$\sigma_{MN}/z^2 \sim \omega/z^2$.
For $k_T^2>0$ this limit is zero and does not contribute. For
$k_T=0$ we have $\varpi =\omega$ and thereby $\sigma_{MN}/z^2 \sim
1/ \omega$ as in four dimensions. Thus, as expected, \textit{only the
zeroth emission mode} $n=0$ contributes in the summation over $n$ of (\ref{DEadd})
in the $\omega\to 0$ limit.

Now, for $k_T=0$  $z=\omega b \psi(\theta) /v$  with
$\psi(\theta)\equiv 1-v \cos\theta$ and
\begin{align}
\tau_{MN}(k) \simeq \frac{ m m' \varkappa_D^2}{4\pi \gamma
V} \!\sum_{l \in \mathbb{Z}^d} \frac{\hat{K}_{1}(|p_T|b)}{\omega^2
b^2 \psi^2(\theta)}\sigma_{MN}\; ,  \qquad p_T^i=\frac{l^i}{R}\,.
\end{align}
For $b\ll R$, which is the case of interest here, the summation over the interaction-KK modes
can be restricted up to modes with $|p_T|b \sim 1$. The result, which can also be obtained by the
usual substitution
of the summation with integration, is
\begin{align}\label{gamoton}
 \tau_{MN}(k) \simeq \frac{\lambda}{\gamma} \frac{2^{d/2}\Gamma(d/2+1)}{\omega^2
b^2 \psi^2(\theta)}\sigma_{MN} \,,
\end{align}
and subsequently, using (\ref{DEadd}), one concludes that
\begin{align}\label{ZFL_ADD}
\left.\frac{dE}{d\omega}\right|_{\omega=0} = \frac{2^{d-2}
\Gamma^2(d/2+1) (\vk^3 m m')^2 }{(2\pi)^{d+4} V b^{2d+2}}\gamma^2
\left[\ln 2\ga +\frac{1}{6} \left(\frac{d}{d+2} -5 \right)\right].
\end{align}
The volume $V$ of the internal space survives in the denominator, because, as explained
above, only the $n=0$ mode of radiation contributes. Of course, for $d=0$ (\ref{ZFL_ADD})
coincides with (\ref{ZFL_4D}).

(d) {\it In ${\mathcal M}_{D>4}$ the frequency spectrum vanishes for $\omega\to 0$}. There are many ways one can
convince himself about this. First, it is obtained by taking the limit $V\to\infty$ of (\ref{ZFL_ADD}). Alternatively,
observe that in higher dimensional Minkowski the relative contribution on the various radiation modes
satisfies $\sum_{n=0}\left/\sum_{n^2>0}\right. \to 0$. Taking in addition into account
that non-zero modes $n$ do not contribute in the limit $\omega\to 0$, one
concludes that $dE/d\omega|_{\omega=0}$ vanishes in higher dimensions.
Finally, the same conclusion is reached by taking the behavior $\tau \sim
1/\omega$, square and multiply by the measure
$\omega^{d+2}$. Again, for $d>0$ the limit of the integrand for $\omega \to 0$ is zero.

\subsection{Quantum constraints}

The above results were obtained in the context of classical theory. However, quantum mechanics restricts the
region of validity of a classical computation.
Thus, to ensure the reliability of the above conclusions, apart from the
condition $b\ll R$, which allows one to replace the summations
over KK modes with integrations, and which will be assumed in the sequel, one has to justify (a) the use of
trajectories for the colliding particles, (b) the expansion around flat space
for the gravitational field, and (c) the use of classical field description of the radiation.
From the analysis of the elastic ultrarelativistic scattering problem \cite{GKST-1} one concludes that the classical
computation around flat space-time of the cross-section is identical to the eikonal approximation of the
quantum answer, as long as the conditions
\begin{equation} \sqrt{s} = 2m\gamma_{\rm cm} \gg M_* \;, \quad b> b_\gamma \equiv r_S \gamma_{\rm cm}^{\; 1/(d+1)}\;, \quad
b<b_c\equiv r_S (r_S/\lambda_B)^{2/d}
\end{equation}
are satisfied\footnote{For colliding particles with equal masses
$\gamma_{\rm cm}^2=(1+\gamma)/2\sim \gamma/2$}. For the radiation problem at hand,
the weak particle-recoil condition due to the emission of
gravitons with characteristic frequency $\omega$ is satisfied if the momenta of
the emitted gravitons are much smaller than the momentum transfer
of the elastic collision. However, experience from analogous computations of the total energy of synchrotron
radiation shows that this condition can be relaxed and replaced, instead, by the
weaker $\om \ll {\mathcal E}_0$, where ${\mathcal E}_0 \simeq m\gamma$ is the total
energy of the colliding particles. When the emitted energy $E$ is
of order ${\mathcal E}_0$, this condition also guarantees a large
number of emitted quanta, and justifies further the description of radiation with a classical field
\footnote{In what follows and in order to simplify notation we
replace the strong inequalities by simple ones.}.

\subsubsection{The $\omega\sim \gamma^2/b$ and $\omega\sim 1/b$ regimes}

$\bullet$ Radiation with characteristic frequency $\omega\sim \gamma^2/b$ does not satisfy the above constraints,
and consequently may not be reliably treated classically in the presence of extra dimensions.
Indeed, for $\omega\sim \gamma^2/b$ the condition
$\omega < m\gamma$ leads to $b>\gamma/m > b_c$, which contradicts the requirement $b<b_c$.
Thus, in the sequel we shall ignore the radiation with characteristic frequency $\gamma^2/b$, even though according to
Table II it is classically dominant for $d\geqslant 2$.

\vspace{0.2cm}

$\bullet$ On the contrary, as explained in \cite{GKST-3} and shown in Table II the radiation emitted in the
soft frequency regime $\omega\lesssim 1/b$, satisfies all the above conditions, but its energy is negligible by phase
space in all dimensions, and will not be discussed further.

\subsubsection{The $\omega \sim \gamma/b$ regime}

According to Table II the dominant radiation in this regime is beamed for $d=0, 1$, almost
isotropic for $d\geqslant 3$,
while for $d=2$ the two components are of comparable importance.

For $\omega\sim \gamma/b$ the quantum condition $\omega <  m\gamma$
reduces to $mb>1$.

Thus, in terms of the rescaled dimensionless parameters
\begin{equation} {\mathcal E}\equiv \frac{\sqrt{s}}{M_*}\;, \quad \mu\equiv \frac{m}{M_*}\;, \quad \beta\equiv bM_*
\end{equation}
a reasonable set of requirements for the validity of the above classical results in this intermediate frequency regime
is \footnote{The condition $\mu\ll 1$ is necessary to justify the use of point-particle, instead of black-hole description
of the colliding particles.}:
\begin{equation} {\mathcal E}\gg 1 \;, \quad  \mu\ll 1 \;, \quad \beta > \left(\frac{{\mathcal E}^2}{\mu}\right)^{1/{(d+1)}}\! \!, \quad
\beta<{\mathcal E}^{2/d}\;, \quad \beta\mu>1\;,
\end{equation}
which are equivalent to
\begin{equation} {\mathcal E}\gg 1 \;, \quad  \mu\ll 1 \;, \quad
\frac{1}{\mu}<\left(\frac{{\mathcal E^2}}{\mu}\right)^{1/(d+1)}<\beta<{\mathcal E}^{2/d}\;,
\label{constraints}
\end{equation}
and can easily be satisfied in all dimensions \footnote{For $d=0$ there is no upper bound on the allowed values
of $\beta$.}.

Thus, we shall next compute the energy emitted in the $\omega\sim
\gamma/b$ characteristic frequency regime \footnote{Incidentally,
this is the frequency regime in which one would a priori expect to
have dominant radiation: Due to the Lorentz squeezing of the field
lines along the direction of relative motion of the colliding
particles, the characteristic duration of the collision in the lab
frame becomes $\tau\sim b/\gamma$. Thus, the characteristic
frequency of the emitted radiation should be $\omega\sim 1/\tau
\sim \gamma/b$.}.

\subsection{Radiation emission with characteristic frequency $\omega\sim \gamma/b$}
\label{zprime}

As explained above, the estimates in Table II were obtained under
the assumption that the frequency integrations could be restricted
to the appropriate regime corresponding to each entry, i.e. from a
fraction of the corresponding characteristic frequency to a few
times that frequency. The quantum conditions discussed above
restrict $\omega$ to $\omega\leqslant \omega_{\rm max}=m\gamma$,
which translates to $z'_{\max}=mb>1$.

\subsubsection{$d\geqslant 3$}

Thus, for $d\geqslant 3$, according to Tables I and II  the
leading contribution to the emitted energy is obtained from
(\ref{Sstress4}) integrated over phase space with $z'$ up to
$z'_{\max}$. But, the fact that the integrand is exponentially
suppressed beyond $z'\sim 1$ , allows one with negligible error
(in the overall coefficient) to extend the $z'$ integration to
$+\infty$. As far as the lower limit
is concerned it is a priori $1/b$, since (\ref{Sstress4}) is valid for $\od > \mathcal{O}(b^{-1})$. However,
since the integrand
in (\ref{Cab}) is regular at $z'=0$ one can shift the lower limit of integration to $0$, the relative
error being of order $1/\gamma^2$.
This reproduces the power of $\gamma$ estimated in
Table II. The purpose of the rest of this subsection is essentially
to obtain the overall coefficient.

Projecting (\ref{Sstress4}) onto $\varepsilon_\pm$ and $\varepsilon_{\rm III}$ one obtains
\begin{align}\label{largeangpol}
& \tau_{\pm}  \simeq   \frac{\lambda}{2 {z'}^2 \gamma \psi}
\left[  (\cos^2 \! \phi \, \cos^2 \! \vartheta-\sin^2  \! \phi
\pm i \sin 2\phi \cos \vartheta) \hat{K}_{d/2+2} (z')  - \sin^2 \!
\vartheta \! \left(\hat{K}_{d/2+1} (z') + \frac{1}{d+2}  {z'}^2
\hat{K}_{d/2}(z') \right)
\right] \nn \\
&  \tau_{\rm III}  \simeq  -  \sqrt{\frac{d}{2(d+2)}}
\frac{\lambda}{ {z'}^2 \gamma \psi} \left[  (1-\cos^2 \! \phi\,
\sin^2 \! \vartheta) \hat{K}_{d/2+2} (z')  - \sin^2 \! \vartheta
\! \left(\hat{K}_{d/2+1} (z') + \frac{1}{d+2}  {z'}^2
\hat{K}_{d/2}(z') \right) \right]\, .
\end{align}
One then according to (\ref{DEMD}) has to square these, multiply with the phase space factor and integrate over
angles and frequencies.

The result of the integration over frequencies takes the form
\begin{align}
\label{CabDab_angles} \frac{dE}{d\Omega_{d+2}} =
\frac{(\varkappa_D^3 m m')^2  \gamma^{d+1}}{16 \, (2 \pi)^{2d+5}
\, b^{3d+3} \,\psi^2} \sum_{a,b=0}^2 C_{ab}^{(d)}
\bar{D}_{ab}^{(d)}(\vartheta, \phi)\,,
\end{align}
where $C_{ab}^{(d)}$ are given by (\ref{Cab})
and
\begin{align}\label{Dab_angles}
 & \bar{D}_{00}^{(d)}(\vartheta, \phi)=\frac{d+1}{(d+2)^3}\sin^4\!\vartheta & & \bar{D}_{01}^{(d)}(\vartheta, \phi)
 =(d+2)  \bar{D}_{00}^{(d)}(\vartheta, \phi)  \nn \\
 & \bar{D}_{02}^{(d)}(\vartheta, \phi)=-\frac{\sin^2\!\vartheta}{d+2}
\left[ \cos^2\! \phi +\frac{2}{d+2}(\cos^2 \! \phi\, \sin^2 \!
\vartheta-1) \right] & & \bar{D}_{11}^{(d)}(\vartheta, \phi)
= (d+2)^2  \bar{D}_{00}^{(d)}(\vartheta, \phi) \nn \\
 & \bar{D}_{12}^{(d)}(\vartheta, \phi)=- (d+2)\bar{D}_{02}^{(d)}(\vartheta, \phi) & & \bar{D}_{22}^{(d)}(\vartheta, \phi)
 =\frac{d+1}{d+2} (1-\cos^2 \! \phi\, \sin^2 \! \vartheta)^2\,.
\end{align}
The values of $C_{ab}^{(d)}$ are to be computed with help of
(\ref{intfreq}).

The integration of (\ref{Dab_angles}, \ref{CabDab_angles}) over
the angles except $\vartheta$ (i.e. over the sphere $S^{d+1}$) is
easily performed using formulae (\ref{angularintegrals}), while
the final integration over $\vartheta$ is done with the help of
(\ref{jj3}) \footnote{ At small angles expressions
(\ref{largeangpol}) are not valid. But as before, if we substitute
integral $\ds \int\limits_{\mathcal{O}(1/\ga)}^{\pi}\!\! d \theta$
by $\ds \int\limits_{0}^{\pi} d \theta$, the relative error is
$\mathcal{O}(1/\ga)$ since the integrand doesn't blow up at
$\vartheta=0+ \mathcal{O}(1/\ga)$. Thus we may apply
(\ref{jj3}).}. Making use also of (\ref{Cab}), one finally obtains
for the emitted energy
\begin{align}
\label{Energy_emit} E = \bar{C}_d \frac{(\varkappa_D^3 m
m')^2}{b^{3d+3} } \gamma^{d+1}\,,
\end{align}
with $\bar{C}_3=1.08\times 10^{-4}$, $\bar{C}_4=4.77\times
10^{-5}$, $\bar{C}_5=4.03\times 10^{-5}$ and $\bar{C}_6=4.98
\times 10^{-5}$.

\subsubsection{$d=0$}

As explained above, the $z'$ integration extends a priori up to
$z'_{\max}\sim mb>1$. It was checked numerically that (a) the
estimate of the cubic power of $\gamma$ given in Table II is
correct, and (b) that the coefficient is insensitive to the upper
limit of integration, varying by a factor of two as one changes
the upper limit of the $z'$ integration between 1 and $\infty$.

Thus, (\ref{Edequal0}) as a reliable estimate.

\subsubsection{$d=1, 2$}

In this case the estimation of the contribution to the emitted energy from the $\omega\sim \gamma/b$
part of the spectrum was done numerically. Specifically, one started with (\ref{Tloc}) and
(\ref{Sstress}) and integrated numerically over $\vartheta$ from 0 to $\pi$ and over $z'$ from 0 to something of order 1.
For a reliable estimate it is enough to integrate up to $z'=1$.

The result for $d=2$ is
\begin{align}\label{Energy_emit6}
 E \simeq 0.04\, \frac{(\varkappa_6^3 m m')^2}{b^{9} } \gamma^{3} \ln\gamma
\end{align}
and for $d=1$
\begin{align}\label{Energy_emit5}
 E \simeq 0.9\times 10^{-4}\, \frac{(\varkappa_5^3 m m')^2}{b^{6} }
 \gamma^{3}\, .
\end{align}


\subsection{Extreme radiation efficiency for $d\geqslant 3 $ and $\omega\sim \gamma/b$}


Consider specifically the case $m=m'$, divide the emitted energy
by the total energy ${\mathcal E}_0 = m+ m\gamma$ of the colliding
particles and use the definition of $r_S$ given in the
Introduction, to write the radiation efficiency $\epsilon\equiv
E/{\mathcal E}_0$ in the form \footnote{Since it is not our
intention to engage in numerology, we ignore in these formulae
inessential d-dependent coefficients, even though they are roughly
of ${\mathcal O}(10^d)$, or the factor $\ln \gamma$, which is
present for $d=2$.} \begin{equation} \epsilon \sim
\left(\frac{r_S}{b}\right)^{3(d+1)} \, \gamma_{\rm cm} \sim
\frac{{\mathcal E}^4}{\mu \,\beta^{\,3(d+1)}}\,,\;\;\; d=0, 1
\label{epsilon012} \end{equation}  and \begin{equation} \epsilon
\sim \left(\frac{r_S}{b}\right)^{3(d+1)} \, \gamma_{\rm cm}^{2d-3}
\sim \frac{{\mathcal E}^{2d}}{\beta^{\,3(d+1)}\mu^{2d-3}}\,,\;\;\;
d\geqslant 2 \,. \label{epsilon3} \end{equation}

It should be pointed out that these expressions differ from the generic formula $\epsilon\sim (r_S/b)^{3(d+1)}
\gamma_{\rm cm}^{2d+1}$ given erroneously in \cite{GKST-PLB}. The difference is due to an error in \cite{GKST-PLB}
related to the extent of destructive interference. Nevertheless, the qualitative conclusions presented in \cite{GKST-PLB}
are still valid. Incidentally, the powers of $\gamma$ in the emitted energy differ from the scalar radiation studied in
\cite{GKST-3} only by the substitution $f\to \varkappa_D m$.

In the parameter range defined in (\ref{constraints}) the radiation efficiency is always much smaller than one
for $d=0, 1, 2$. Indeed, consider the case $d=0$. The maximum $\epsilon$ is obtained for the minimum allowed
value $\beta \sim ({\mathcal E}^2/\mu)^{1/(d+1)}$ of the impact parameter and is equal to
\begin{equation} \epsilon_{\rm max} (d=0) \sim \left(\frac{\mu}{{\mathcal E}}\right)^2 \ll 1\,.
\end{equation}
Thus, in $d=0$ a small fraction of the available energy is emitted in gravitational radiation
and vanishes in the massless limit.
However, one can convince her/himself that for $d\geqslant 3$ the efficiency can take
values arbitrarily close to one or even become {\it much greater than one} for a wide range of parameters obeying the
{\it classicality conditions} (\ref{constraints}) stated above.
For instance, in $d=4$ the minimum value of $\epsilon$ is obtained
for $\beta \sim {\mathcal E}^{2/d}$ and is
\begin{equation} \epsilon_{\rm min} (d=4) \sim \frac{\sqrt{{\mathcal E}}}{\mu^5} \gg 1\;.
\end{equation}

Ditto for dimensions $d > 4$. This is a priori a puzzling result.
It leads for $d\geqslant 3$ to unphysical, i.e. greater than one,
values of the radiation efficiency, and actually blows-up in the
massless limit ($\mu\to 0$). In relation to this we would like to
offer the following comments: (a) The massless limit is a very
singular limit in our approach. The characteristic frequencies and
angles all become singular in that limit. In particular, since in
that limit $b_c / b_\gamma \sim ({\mathcal E}^{2/d} \mu)^{1/(d+1)}
\to 0$, it is impossible to satisfy (\ref{constraints}), even if
one ignores the ``quantum" constraint $\beta \mu > 1$. Thus, the
massless limit lies outside the domain of validity of our
approximation. (b) Weinberg has shown \cite{W} that in $d=0$ there
is no infrared divergence in the {\it soft graviton} emission rate
from massless colliding particles. This is in agreement with our
results not only for the soft component in $d=0$, but also for
soft as well as high frequency emission in $d=0, 1, 2$. (c) The
explicit computation presented here is rather involved, but, as
explained, the powers of $\gamma$ in the emitted energy are
determined on general grounds, based on properties of Macdonald
functions, number of KK interaction and emission modes and
dimensionality of phase space. (d) One may suspect that some
classicality condition is missed, which might render the above
formulae non-applicable. But the constraints used in
(\ref{constraints}) are quite straightforward and based on
well-known physics. So, (e) it may well be the case, that the
above results are reliable and the interpretation of their
puzzling features is that in ultra-planckian particle collisions
there is a lot of energy emitted in gravitational radiation. This
will lead to strong radiation damping, which should be taken into
account in the study of such ultra-planckian scattering processes.

\section{Conclusions}

A detailed study was presented of {\it classical gravitational radiation} emitted in ultra-relativistic collisions of
massive point-particles interacting gravitationally.
The space-time was assumed to have an arbitrary number of toroidal or non-compact extra dimensions and
the post-linear approximation scheme of General Relativity was employed for the computation.
The angular and frequency distributions of radiation, as well as the total emitted energy were studied in detail
to leading ultra-relativistic order.

Three characteristic frequency regimes ($1/b$, $\gamma/b$ and $\gamma^2/b$)
of the emitted radiation were identified and the characteristics of the dominant contribution was determined
in various dimensions.

In particular, in any number of dimensions the soft component of radiation is mainly due to the scattered
particles, with negligible contribution coming from the cubic graviton interaction term \footnote{This is a well
known fact, verified easily also in the context of Feynman diagram infrared graviton summation.}.
However, in contrast
to the four-dimensional case, in any
number of extra dimensions $d>0$ the frequency
spectrum of the emitted radiation vanishes as $\omega\to 0$ and the total emitted
energy in soft gravitons is negligible.

Also, contrary to what happens with the soft radiation emission, the cubic graviton interaction and the scattered
particles themselves are equally important as sources of radiation with high frequency. In fact it was shown that in any
dimension they lead to partial cancellation ({\it destructive interference}) of the total beamed radiation
amplitude in the high frequency domain, as a result of which the emitted energy in the $\gamma^2/b$
frequency regime is reduced by two powers of the Lorentz factor $\gamma$ in the Lab frame.

The relevance of the classical analysis to the full {\it quantum radiation} problem was also discussed.
The {\it classicality conditions}, necessary for the classical treatment to be a good approximation to the full
quantum problem were derived and the radiation efficiency $\epsilon$,
i.e. the fraction of the initial energy which is emitted
in gravitational radiation, was computed for parameter values inside
the region of validity of our classical computation. Although for $d=0, 1, 2$, $\epsilon$ is smaller than one
in the whole allowed range of parameters, it can be arbitrarily close to one or even exceed unity in $d\geqslant 3$.
One possible interpretation of this ``unphysical" result is that indeed in such ultra-planckian collisions
there is a lot of gravitational radiation, which will lead to strong damping, which should be included in a reliable
treatment of the scattering process.

However, the implementation of this interpretation, the proper treatment of the massless limit, the
extension of the region of validity to impact parameters outside the range (\ref{constraints}) and the
comparison with quantum results based, for example, on string theory, are currently under investigation and, hopefully,
will be the subject of a future publication.

\vspace{0.3cm}

\section*{Acknowledgements}
We are grateful to Dr. Georgios Kofinas for useful discussions and
to Mr. Yiannis Constantinou for help with the numerical work. This
work was supported in part by EU grants PERG07-GA-2010-268246, the
EU program ``Thales'' ESF/NSRF 2007-2013 and grant 11-02-01371-a
of RFBR. It has also been co-financed by the European Union
(European Social Fund, ESF) and Greek national funds through the
Operational Program ``Education and Lifelong Learning'' of the
National Strategic Reference Framework (NSRF) under
 ``Funding of proposals that have received a positive evaluation in the 3rd and 4th Call of ERC Grant Schemes''.
DG and PS are grateful to the Department of Physics of the University
of Crete for its hospitality in various stages of this work. PS is also grateful to the DAAD service for co-funding and
to the LMU, Munich, for its hospitality during the final stage of the work.

\appendix

\section{Definitions and variations}
\label{app1}

\subsection{Perturbation theory variation over gravitational constant in flat background}
We use the notation of Weinberg \cite{Weinberg}.
\begin{align}
\label{elemvars}  & g^{(1)}_{MN}=h_{MN} &\qquad&
\Gamma^{(1)R}_{MN}=(h^{R}_{M \cd N}+h^{R}_{N\cd
M}-h_{MN}^{\quad\cd R})/2   \nn\\&
  \Gamma^{(1)N}_{MN}=h_{\cd M}/2 &\qquad&
  \Gamma^{(1)M}_{NR}\eta^{N
R}  =0\, .
\end{align}
The definition of Riemann and Ricci tensors:
\begin{align}\label{elemvars0}
R^{M}{}_{\!\!NLR}=\Gamma^{M}_{NR,
L}-\Gamma^{M}_{NL,R}+\Gamma^{M}_{SL}\Gamma^{S}_{NR}-\Gamma^{M}_{SR}\Gamma^{S}_{NL}\,,
\qquad
 R_{MN} \equiv R^{L}{}_{MLN}\, .
\end{align}
Thus
\begin{align}\label{elemvars1}
&   R_{MN}^{(1)}=\frac{1}{2}(h^{R}_{N; \, MR}+h^{R}_{M\cd
NR}-\Box\, h_{MN}-h_{\cd MN}) =- \frac{1}{2} \Box\,
h_{MN} \nn\\
&   R^{(1)}=-\Box\, h+h_{MN}^{\quad\cd MN}-R^{MN}h_{MN}=-
\frac{1}{2}
\Box\, h \nn\\
& G_{MN}^{(1)}=\frac{1}{2}(-\Box\,\psi_{MN}-\eta_{MN}\xi_{L}^{\;
\cd L}+\xi_{M \cd N}+\xi_{N \cd M})=-\frac{1}{2}\Box\,\psi_{MN}
\qquad \xi_{M}=\partial^{N}\psi_{MN}=0\, .
\end{align}
The second-order variations:
\begin{align}\label{elemvars2}
& \Gamma^{(2)M}_{LR}=-\frac{1}{2}h^{MN}(h_{LN\cd R}+
h_{RN\cd L}-h_{LR\cd N})  \nn\\
 &
  \Gamma^{(2)L}_{ML}=
-\frac{1}{2}h^{LR}h_{LR\cd M}  \qquad
\Gamma^{(2) M}_{ N R} \eta^{ N R} =0 \nn\\
 &  R_{MN}^{(2)} =-\frac{1}{2}\left({h}_M^{A \cd B}(h_{NB \cd A} - h_{NA
\cd B})-\frac{1}{2} h^{AB}_{\quad \cd M} h_{AB \cd N}
+h^{AB}(h_{MA \cd NB}+ h_{NA \cd MB}- h_{AB \cd MN}- h_{MN \cd
AB})  \right) \nn \\
 &  R^{(2)}=(g^{LR} R_{LR})^{(2)}=
 -h^{LR}\,   R_{LR}^{(1)}+ \eta^{LR} \,   R_{LR}^{(2)}=
 \frac{1}{2} \left( 2h^{AB}\Box\,
h_{AB}-h_{AB \cd L} h^{AL \cd B}+\frac{3}{2} h_{AB \cd L} h^{AB\cd
L}\right)\nn \\
  & G_{MN}^{(2)}=
  R^{(2)}_{MN}-\frac{1}{2}\eta_{MN}\,
  R^{(2)}-\frac{1}{2}h_{MN}\,
  R^{(1)}\, .
\end{align}
As a result:
\begin{align}\label{natag}
   \!\!\!\!\! S_{MN} = -  G_{MN}^{(2)}  =\frac{1}{2} & \left[h^{AB}(h_{MA \cd NB}+ h_{NA \cd MB}-
h_{AB \cd MN}- h_{MN  \cd AB}) -\frac{1}{2} h^{AB}_{\quad \cd M}
h_{AB \cd
N}-\frac{1}{2}h_{MN}\Box\, h  + \right.\nn\\
& \left. +{h}_M^{A \cd B}(h_{NB \cd A} - h_{NA \cd B}) +
\frac{1}{2}\eta_{MN}\left(2h^{AB}\Box\, h_{AB}-h_{AB \cd L} h^{AL
\cd B}+\frac{3}{2} h_{AB \cd L} h^{AB\cd L}\right)\right].
\end{align}

\subsection{Useful kinematical formulae}\label{formulae}

The angles in the formulae below are defined in Fig.\ref{branepic}.
\begin{align}\label{kin_form}
&u^{\mu}\!\equiv \ga(1, 0, 0, v) \,, \quad u'^{\mu}\equiv (1, 0,
0, 0)\,, \quad
\psi\equiv 1-v\cos\theta\cos\alpha=1-v\cos\vartheta \,,  \quad {\uln}^i=\frac{n^i}{R}\nonumber \\
& \varpi\equiv |\mathbf{k}|\, ,\quad  \cos
\alpha=\frac{\varpi}{\omega}\, ,\quad z'\!=\frac{(ku')b}{\ga
v}\!=\!\frac{\om b}{\ga v }\,, \;\; z\!=\!\frac{(ku)b}{\ga
v}\!=\!\frac{\om b}{v }\, \psi=z' \ga \psi\,, \;\;
z_l^2\!\equiv \!z^2+b^2 {\ull}^2\, , \quad  {\ull}^i=\frac{l^i}{R} \nonumber \\
&  \xi^2\equiv 2\ga z z'-z^2-{z'}^2
= \varpi^2 b^2 \sin^2\theta + b^2 \uln^2 = (\om  b \sin \vartheta)^2=(\ga v z' \sin\vartheta)^2 \nonumber \\
&-(kb)=\xi \cos\phi = \gamma z' v \sin \vartheta \cos\phi =\gamma
z' v\cos \alpha
\sin \theta \cos \varphi = \varpi b \sin \theta \cos \varphi        \nonumber \\
&\beta\equiv \ga z z' - z^2 = \frac{ \varpi^2 b^2 \cos\vartheta
(1-v\cos\vartheta)}{v \cos^2\alpha} =\ga^2 {z'}^2 \psi (1-\psi)\,
.
\end{align}

Next we compute non-vanishing products of polarization vectors:
Using (\ref{u}), (\ref{kam}), (\ref{kamu}) and these definitions
one finds the following scalar products
\begin{equation}(ku)=\ga\left(\sqrt{\varpi^2+\uln^2}-\varpi
v\cos\theta\right)\, , \qquad (ku')=\sqrt{\varpi^2+\uln^2}\,,
\qquad (uu')=\ga \, .
\end{equation}
To calculate the scalar product $( b e_2)$ one observes that
$\epsilon^{x z 0 \tau 3...{D-2}} b_x u_{z} u'_{0} k_{\tau}
e_{3}...e_{D-2}=-b \gamma v k_\bot ,$ where $k_\bot$ is the length
of the projection of $k^M$ onto the subspace orthogonal to
$t,x,z$, since $k$ is transverse to all $e_{\alpha}$'s and the
form $\epsilon^{\tau 3...{D-2}}k_{\tau} e_{3}...e_{D-2}$,
remaining after factoring out the $t,x,z$ subspace, represents the
volume of the rectangle. Thus
\begin{equation}
k_\bot=\sqrt{\varpi^2\sin^2\theta\sin^2\!\ffi+\uln^2}=\od \sin
\phi\, ,
\end{equation}
and one finds
\begin{equation}
(b e_2)=-b\;\left(\fr{\varpi^2\sin^2\theta\sin^2\ffi+\uln^2}
{\varpi^2\sin^2\theta+\uln^2}\right)^{1/2}=-b\;\left({\fr{ \cos^2
\alpha \sin^2\theta\sin^2\ffi +\sin^2 \alpha} {\cos^2 \alpha
\sin^2\theta+\sin^2 \alpha}}\right)^{1/2} =-b\sin\phi\, .
\end{equation}
Evaluation of the scalar product $(e_1 b)$ is straightforward and
leads to
\begin{equation}
(b e_1)=-\fr{b}2\;\fr{\varpi^2 \sin
2\theta\cos\ffi}{\sqrt{(\varpi^2\sin^2\theta+\uln^2)(
 \varpi^2 +\uln^2)}}=-\frac{b}{v} \cos \phi
\left(1-\frac{(ku)}{\gamma (ku')}\right)=\frac{(k b)}{(k u')}\ctg
\vartheta \, .\end{equation}
 Finally, the remaining non-vanishing
 scalar product is
 \begin{equation}
 (e_1 u)=\ga v \left(\frac{ {\varpi^2\sin^2\theta +\uln^2 }}
 { {\varpi^2 +\uln^2}}\right)^{1/2}=\ga v\sqrt{{\cos^2 \alpha \sin^2\theta+\sin^2 \alpha} }=\gamma v \sin\vartheta\, .
 \end{equation}

\section{Momentum integrals}
\label{app2}

\subsection{Basic scalar integral}
The local integrals for massless and massive modes are calculated
in \cite{GKST-2}. The main integral for stress (also for massive
modes) is calculated in \cite{GKST-3}.
\begin{align} J(k)\equiv \frac{1}{V}
\sum_{l }J^{nl}(k)\,, \qquad J^{nl}(k)= \int  \,d^4 p
\frac{\delta(pu')\delta(ku-pu) e^{-i(pb)}}{ (p^2-\ull^2)\, [
(k-p)^2 - (\uln^i - \ull^i)^2] } \,.
 \end{align}
$\omega \equiv k^0=\sqrt{\mathbf{k}^2+{\uln}^2}$ and $\mathbf{k}$
is a $3-$dimensional vector lying on the $3-$brane, while
${\uln}^i=n^i/R$ and ${\ull}^i=l^i/R$ with integers $\{n^i\},
\{l^i\}$, are $d-$dimensional discrete vectors, corresponding to
the emission and interaction modes, respectively.

Using Feynman parametrization $J^{nl}$ takes the form:
\begin{align}
J^{nl}=\int\limits_0^1  dx \, e^{- i (kb)x} \int d^4 p \frac{
\delta[(pu')+ (ku') x]
\delta[(pu)-(1-x)(ku)]\,\e^{-i(pb)}}{[p^2-({\uln}
x-{\ull})^2]^2}\,.
\end{align}
Integrating over $p^0$ and splitting $\mathbf{p}$ into the
longitudinal $p_{||}$ (with respect to $\mathbf{u}$) and transversal $\mathbf{p_{\bot}}$ parts,
integrate over $p_{||}$. The corresponding covariant splitting of $b^{\mu}$ on temporal, longitudinal and
transversal parts  reads
$$b^{\mu}=(bu'){u'}^{\mu}-\frac{\gamma(bu')-(bu)}{\gamma^2 v^2}(\gamma {u'}^{\mu} - u^{\mu} )+b_{\bot}^{\mu}$$
and denote scalar impact parameter $b$ as
\begin{align}\label{b_tr}
b \equiv
\sqrt{-b_{\bot}^2}=\left(-b^2-\frac{[(bu)u'-(bu')u]^2}{\gamma^2
v^2}  \right)^{1/2}\!.
\end{align}

Then, introducing in $\mathbf{p_{\bot}}$
the spherical coordinates
$d^{2}\mathbf{p_{\bot}}=|\mathbf{p}_\perp|d\Omega_{1}
d|\mathbf{p}_{\bot}|$ and integrating first over the angles and
then over $|\mathbf{p}_{\bot}|$, one obtains
\begin{align}
\label{hhh6} & J^{nl} =\frac{\pi b^2}{\gamma v} \int\limits_0^1
dx\,e^{-i (Nb)}\, \hat{K}_{-1}(\zeta_{nl})\,,
\end{align}
with
\begin{align}
\label{hhh5} &
\zeta_{nl}^2(x)= {z'}^2 x^2 + 2\ga z z' x(1-x) + z^2
(1-x)^2 +b^2({\uln} x-{\ull})^2\nn \\
& N_\mu= \frac{1}{\gamma^2 v^2}\left[
x(ku')u'_{\mu}-(1-x)(ku)u_{\mu}\right]-\frac{1}{\gamma v^2}\left[
x(ku')u_{\mu}-(1-x)(ku)u'_{\mu} \right]+x k_{\mu}
 \,.
\end{align}
Sum up over $l'$s with rule (\ref{sum2int}) to get
\begin{align}
\label{cucu1} J^n(k) =\Lambda_d \int\limits_0^1 dx\,e^{-i(Nb)}\,
\hat{K}_{d/2-1}(\zeta_n)  ; \qquad \Lambda_d\equiv \frac{\pi
b^{2-d}}{(2\pi)^{d/2} \gamma v}\,.
\end{align}
with
\begin{align}
\label{cucu2}
\zeta_n^2(x)= {z'}^2 x^2 + 2\ga z z' x(1-x) + z^2
(1-x)^2  \,; \qquad  \zeta_{n}(0)=z\,,\;\; \zeta_{n}(1)=z' \,.
\end{align}
Finally, fix $(bu')=(bu)=0$ to get
\begin{align}
\label{cucu1_5}
J^n(k) =\Lambda_d \int\limits_0^1
dx\,e^{-i(kb)x}\, \hat{K}_{d/2-1}(\zeta_n)  \,.
\end{align}
Thus
\begin{align}
\label{cucu2_1} J^n(k) =J^{[z]}+J^{[z']}.
\end{align}
In the frequency-angular region $\varrho \sim \od b \vartheta \gg
1$ $J^{[z]}$ and $J^{[z']}$ are given by the series over small
$1/\varrho^2$:
\begin{align}
&\label{J0} {J}^{[z]}  =  \frac{\Lambda_{d}}{a^2 \coa^2} \left(\!
\beta \hat{K}_{d/2}(z)-i
 (kb)\hat{K}_{d/2+1}(z)  -\frac{(d+1)\beta}{a^2}
\hat{K}_{d/2+1}(z) \! + \frac{\beta \sin^2\! \!\phi
}{a^2}\hat{K}_{d/2+2}(z) \right)+R_{z}
\\
&\label{J1} {J}^{[z']} = \Lambda_{d} \, \e^{-i (k b)}
\left(\frac{\coa^2-\beta}{\coa^2 a^2}\,
\hat{K}_{d/2}(z')-i\frac{\cos\!\phi}{a^2 \,\xi} \,
\hat{K}_{d/2+1}(z')\right)+R_{z'}\, ,
\end{align}
where $$a\equiv \sqrt{\frac{\beta^2}{\xi^2}+z^2}=\frac{z}{\sin
\vartheta}\, .$$

Thus $J^n(k)$ can be split in two parts with drastically different
spectral-angular behavior each.

\subsection{Vectorial and tensorial integrals}

Vectorial integral is defined by
\begin{align} J^n_{M}(k)\equiv \frac{1}{V}
\sum_{l }J^{nl}_{M}(k)\,, \qquad J^{nl}_{\mu}(k)= \int  \,d^4 p \,
\frac{\delta(pu')\delta(ku-pu) e^{-i(pb)}}{ (p^2-\ull^2)\, [
(k-p)^2 - (\uln^i - \ull^i)^2] }\, p_{M} \,.
 \end{align}
Thus
$$J^n_{M}(k) =i \frac{\partial J}{\partial b^{M}}\,.$$
Substituting $J^n(k)$ in the form (\ref{cucu1}) and
differentiating, one gets taking (\ref{b_tr}) into account
\begin{align}\label{susu0}
 J^n_{M} =\Lambda_d \int\limits_0^1 dx\,e^{-i(Nb)} \left[
N_{M}\hat{K}_{d/2-1}(\zeta_{n})+i \frac{ \hat{b}_{M}}{b^2}
\hat{K}_{d/2}(\zeta_{n})\right], \end{align} with
\begin{align}
\hat{b}_{M}\equiv {b}_{M}-\frac{(bu) u_{M} +(bu')
{u'}_{M}-\gamma[(bu) {u'}_{M} +(bu')u_{M} ]}{\gamma^2 v^2}\, ,
\qquad N_M=(N_{\mu}; x  {k_T}^i)\, ,
\end{align}
where it is convenient to use the
following properties of hatted Macdonalds:
$$\hat{K}'_{\pm n}(x) =-x \hat{K}_{\pm n-1}(x) \, .$$
In special frame
\begin{align}
\label{susu0_5}
 J^n_{M} =\Lambda_d \int\limits_0^1 dx\,e^{-i(kb)x} \left[
N_{M}\hat{K}_{d/2-1}(\zeta_{n})+i \frac{ b_{M}}{b^2}
\hat{K}_{d/2}(\zeta_{n})\right]\,.
\end{align}
Some useful products
\begin{align} \arraycolsep=0.7cm
\begin{array}{lll}
 N \cdot u=(ku) &  N \cdot u' =J^n
\cdot u'=0 & \ds N\cdot k=\frac{1}{b^2}[x\left(z^2+z'^2-2z
z'\gamma\right)+\left(\gamma z z'-z^2 \right)] \\
 J^n  \cdot u =(ku)J^n & N \cdot b = x(kb) & \ds N^2=\frac{1}{b^2}\left[x^2\left(z^2+z'^2-2z z'\gamma\
\right)-z^2\right].
\end{array}\nn
\end{align}
Tensorial integral is defined
\begin{align} J_{MN}(k)\equiv \frac{1}{V}
\sum_{l }J^{nl}_{MN}(k)\,, \qquad J^{nl}_{MN}(k)= \int  \,d^4 p \,
\frac{\delta(pu')\delta(ku-pu) e^{-i(pb)}}{ (p^2-\ull^2)\, [
(k-p)^2 - (\uln^i - \ull^i)^2] }\, p_{M}  p_{N}\,.
 \end{align}
Thus
$$J^n_{MN}(k) =i \frac{\partial J^n_{M}}{\partial b^{N}}.$$
Substituting $J^n_M$ in the form (\ref{susu0}) and
differentiating, we have
\begin{align}
\label{iij1} J^n_{MN} = \frac{\Lambda_d}{ b^2} \int\limits_0^1
dx\,e^{-i(kb)x} &  \left[b^2 N_{M} N_{N} \hat{K}_{d/2-1}(
\zeta_n)+ 2iN_{(M} b_{N)} \hat{K}_{d/2}( \zeta_n)-
\frac{b_{M}b_{N}}{b^2}\hat{K}_{d/2+1}( \zeta_n)- \right.\nn \\
& \left. -\left( \eta_{MN}+\frac{1}{v^2
\gamma^2}\left[u_{M}u_{N}+u'_{M}u'_{N}-2\gamma
u_{(M}u'_{N)}\right] \right)\!\hat{K}_{d/2}( \zeta_n)\right].
\end{align}
Some products:
\begin{align}\label{iij2}
 J^n_{MN}u^{N} =(ku)J^n_{M} \quad\qquad J^n_{MN}u'^{N} =0 \quad\qquad
 J_{MN}u^{M}u^{N} =(ku)^2 J^n.
\end{align}
 Here we see that integrals $J^n_{M}, \; J^n_{MN}$ represent the
superposition of some scalar integrals of the type
\begin{align}
\label{iii4}
 {J}_{(\sigma, \tau)} \equiv \Lambda_d \int\limits_0^1 x^{\sigma}\, e^{-i(kb)x}
 \hat{K}_{d/2+\tau}(\zeta_n) \, dx\,,
\end{align}
(with $\sigma=0,1,2$): for instance
\begin{align}
\label{iii2a}
  J^n_{M} =\frac{1}{v b}\left[ \left( \frac{1}{\gamma}\left[ z'
u'_{M}+z u_{M}\right]- \left[ z' u_{M}+z u'_{M} \right]+v b
k_{M}\right) {J}_{(1,-1)}
+z\left(u'_{M}-\frac{1}{\gamma}u_{M}\right){J}_{(0,-1)}+i \frac{v
b_{M}}{b}{J}_{(0,0)}\right].
\end{align}
Two integrals  ${J}_{(0,-1)}$ and ${J}_{(0,0)}$  (i.e. with
$\sigma=0$) are of the type of the basic scalar integral and
thereby known, while the derived one ${J}_{(1,-1)}$ is new. The
computation of integrals with $\sigma=1,2$ represents the goal of
next subsection.

\subsection{Derived integrals for stresses}
\label{derintsS}
Now consider (\ref{iii4}) with $\sigma=1,\, \tau=-1:$
\begin{align}\label{iii4d}
 {J}_{(1,-1)}=\Lambda_d \int\limits_0^1 x\, e^{-i(kb)x}
 \hat{K}_{d/2-1}(\zeta_n) \, dx\,,
\end{align}

If we'd know $J(k)$  exactly (\ref{iii4d}) can be calculated by
differentiation:
$${J}_{(1,-1)}=-\frac{i}{\xi}\frac{\partial {J}_{(0,-1)}}{\partial
\cos \! \phi},$$ Our strategy consists in the reduction of $
{J}_{(\sigma, \tau)}$ into the superposition of ${J}_{(0, \tau')}$
(with some $\tau'$s), for which we already have calculated
expressions (\ref{J0}, \ref{J1}). Note that for higher $\tau$ we
keep $\Lambda_d$ and shift index of Macdonalds in (\ref{J0},
\ref{J1}): thereby (\ref{J0}, \ref{J1}) may be generalized into
\begin{align} \label{J01pr}
& {J}^{[z]}_{(0,\tau-1)}\!=\! \frac{\Lambda_{d}}{a^2 \coa^2}
\left(\! \beta \hat{K}_{d/2+\tau}(z)-i
 (kb)\hat{K}_{d/2+\tau+1}(z)  -\frac{(d+2\tau+1)\beta}{a^2}
\hat{K}_{d/2+\tau+1}(z) \! + \frac{\beta \sin^2 \!\phi
}{a^2}\hat{K}_{d/2+\tau+2}(z) \right)\nn
\\
& {J}^{[z']}_{(0,\tau-1)} \simeq \frac{\Lambda_{d} \, \e^{-i (k
b)}}{a^2 \,\xi^2} \left[(\coa^2-\beta)\,
\hat{K}_{d/2+\tau}(z')-i\xi \cos\phi \,
\hat{K}_{d/2+\tau+1}(z')\right]\, .
\end{align}

 \vspace{0.3cm}

 Representing $x=[x-\beta/\coa^2]+\beta/\coa^2$, we
have
\begin{align}\label{iii5}
{J}_{(1,-1)}=\frac{\beta}{\coa^2}{J}_{(0,-1)}+\frac{\Lambda_d}{\coa}\int\limits_0^1
r\, e^{-i(k b)x}
 \hat{K}_{d/2-1}(\zeta_n) \,
 dx=\frac{\beta}{\coa^2}{J}_{(0,-1)}+\frac{\Lambda_d}{\coa}\bar{J},
\end{align}
with notation
\begin{align}\label{iii6}
\bar{J}=\int\limits_0^1 dx\,r(x) e^{-i(k b)x} \hat{K}_{d/2-1}(
\zeta_n)\, .
\end{align}
Representing \cite{GR, Proudn}
\begin{align}\label{hhh11}
\hat{K}_{\nu-\mu-1}\left( \sqrt{a^2-r^2}\right)= r^{-\mu}a^\nu
\int \limits_0^\infty
 \frac{ y^{\mu+1}I_\mu(r y)}{(y^2+1)^{\nu/2}}K_\nu
 (a \sqrt{y^2+1})dy\, ,
\end{align}
we have
\begin{align}\label{iii6b}
\bar{J}=  \int\limits_0^\infty dy \,
 \frac{a^\nu y^{\mu+1}}{(y^2+1)^{\nu/2}}K_\nu
 (a\sqrt{y^2+1}) \,
\int\limits_0^1 dx\, e^{- i(k b)x} r^{1-\mu}I_\mu(r y)\, .
\end{align}
Fix $\mu=1/2$, then $\nu=(d+1)/2=(D-3)/2>-1$, and
\begin{align}\label{iii7}
r^{1/2}I_{1/2}(r
y)y^{3/2}=y\frac{\sqrt{2}}{\sqrt{\pi}}\sinh(ry)\,.
\end{align}
Taking the internal integral over $x$
\begin{align*}
\int\limits_0^1 dx\, e^{-i(kb)x}\sinh(\coa[x- \beta/\coa^2] y)=
\sum_{j=0,1}(-1)^{j-1} e^{-i(kb)j}\frac{y \coa \cosh (\coa d_j
y)+i(kb)\sinh (\coa d_j y)}{\coa^2 y^2+(kb)^2}\, ,
\end{align*}
with
\begin{align}
d_j=j-\beta/\coa^2=\delta_{1j}-\beta/\coa^2 \qquad j=0,1 \, , \nn
\end{align}
we arrive at
\begin{align}\label{iii9}
\bar{J}=\frac{\sqrt{2}}{\sqrt{\pi}}  a^{(d+1)/2}
\sum_{j=0,1}(-1)^{j-1} e^{-i(kb)j} \int\limits_0^\infty dy \,
 \frac{K_{(d+1)/2}
 (a\sqrt{y^2+1})}{(y^2+1)^{\nu/2}}\frac{y^2 \coa\cosh \coa
d_j y+i y (kb)\sinh \coa d_j y}{\coa^2 y^2+(kb)^2}\,.
\end{align}
Substituting $(kb)=-\coa \cos \phi$ (\ref{kin_form}) to get
\begin{align}
\bar{J}=\frac{\sqrt{2}}{\sqrt{\pi}}  \frac{a^\nu}{\coa}
\sum_{j=0,1}(-1)^{j-1} e^{-i(k b)j} \int\limits_0^\infty dy \,
 \frac{K_{(d+1)/2}
 (a\sqrt{y^2+1})}{(y^2+1)^{\nu/2}}\frac{y^2 \cosh \coa
d_j y- i y \cos \phi\sinh \coa d_j y}{  y^2+\cos^2 \!\phi}\, . \nn
\end{align}
In the integrand  numerator add and subtract $\cos^2 \!\phi$ to
$y^2$:
\begin{align}
\bar{J}=\frac{\sqrt{2}}{\sqrt{\pi}}  \frac{a^{(d+1)/2}}{\coa}
\!\!\sum_{j=0,1}\!(-1)^{j-1} e^{-i(kb)j} \!\!\int\limits_0^\infty
\! dy \,
 \frac{K_{(d+1)/2}
 (u \sqrt{y^2+1})}{(y^2+1)^{(d+1)/4}}\! \left[\cosh \coa
d_j y- \frac{\cos^2 \!\phi \cosh \coa d_j y+i y \cos \phi\sinh
\coa d_j y}{  y^2\!+\!\cos^2 \!\phi}\right] \nn
\end{align}
and then, integrating $\cosh \coa d_j y$ in a bracket with help of
(\ref{hhh11}) and comparing the remainder with
\begin{align} {J}_{(0,0)} = \frac{2^{1/2}
a^{(d+1)}}{\pi^{1/2}} \, \frac{\Lambda_d}{\xi} \,\sum_{j=0,1}
\!(-1)^{j+1} e^{-ij (k  b)} \! \!\int\limits_0^\infty \! dy \,
\hat{K}_{-(d+1)/2} \! \left(a \sqrt{y^2+1}\right)\frac{y
\sinh(\coa\delta_j y)-i\cos\phi \cosh(\coa \delta_j
y)}{y^2+\cos^2\!\phi} \nn
\end{align}
\cite[eqn.(3.28)]{GKST-3}, one concludes
\begin{align} \bar{J}= \frac{1}{\xi}
\left[e^{-i  (k  b)} \hat{K}_{d/2}(z') - \hat{K}_{d/2}(z)\right]-i
\frac{\cos   \phi}{\Lambda_d} \,{J}_{(0,0)}\, .
\end{align}
Substituting into (\ref{iii5}) one gets the recurrence relation:
\begin{align}
{J}_{(1,-1)}=\frac{\beta}{\coa^2}{J}_{(0,-1)}+
\frac{\Lambda_d}{\coa^2} \left[e^{-i  (k  b)} \hat{K}_{d/2}(z') -
\hat{K}_{d/2}(z)\right]-\frac{i \cos \phi }{ \coa}\,{J}_{(0,0)}\,
.
\end{align}
The property is hold for all indices ($\tau\geqslant -1$), thus we
can generalize:
\begin{align}\label{reduct}
 {J}_{(1,\tau)}=\frac{\beta}{\coa^2} {J}_{(0,\tau)}+
\frac{\Lambda_d}{\coa^2} \left[e^{i \coa \cos \phi}
\hat{K}_{d/2+\tau+1}(z') - \hat{K}_{d/2+\tau+1}(z)\right]-\frac{i
\cos   \phi }{
\coa}\,{J}_{(0,\tau+1)}=-\frac{i}{\coa}\frac{\partial
 {J}_{(0,\tau)} }{\partial(\cos  \phi)}\, .
\end{align}
Splitting it onto $z-$ and $z'-$ parts gives
\begin{align}\label{reduct02}
&{J}_{(1,\tau)}^{[z]}=\frac{\beta}{\coa^2} {J}_{(0,\tau)}^{[z]}-
\frac{\Lambda_d}{\coa^2} \hat{K}_{d/2+\tau+1}(z) -\frac{i \cos
\phi }{ \coa}\,{J}_{(0,\tau+1)}^{[z]}\, \nn  \\
&{J}_{(1,\tau)}^{[z']}=\frac{\beta}{\coa^2} {J}_{(0,\tau)}^{[z']}+
\frac{\Lambda_d}{\coa^2}  e^{i \coa \cos \phi}
\hat{K}_{d/2+\tau+1}(z')  -\frac{i \cos  \phi
}{\coa}\,{J}_{(0,\tau+1)}^{[z']}\, .
\end{align}

Differentiating (\ref{reduct}) over $\cos \phi$, one gets
\begin{align}\label{reduct1}
{J}_{(2,\tau)}=\frac{\beta}{\coa^2}{J}_{(1,\tau)}+
\frac{\Lambda_d}{\coa^2} e^{i \coa \cos \phi}
\hat{K}_{d/2+\tau+1}(z') -\frac{1}{\coa^2} {J}_{(0,\tau+1)}-
\frac{i \cos \phi }{ \coa}\,{ J}_{(1,\tau+1)}\, .
\end{align}
Substituting (\ref{reduct})
\begin{align}
\label{reduct2}
 {J}_{(2,\tau)}=&\frac{\beta^2}{\coa^4} {J}_{(0,\tau)}+\left.\frac{\Lambda_d \beta}{\coa^2}
e^{i  x \coa \cos \phi} \hat{K}_{d/2+\tau+1}(
\zeta_n)\right|_0^1-\frac{2i \beta \cos \!\phi }{ \coa^3}\,{
J}_{(0,\tau+1)} + \frac{\Lambda_d}{\coa^2} e^{i \coa \cos \phi}
\hat{K}_{d/2+\tau+1}(z')-\nn \\& -\frac{1}{\coa^2}
{J}_{(0,\tau+1)}- \left.  \frac{i \Lambda_d\cos \!\phi }{
\coa^3}\, e^{i x \coa \cos \! \phi} \hat{K}_{d/2+\tau+2}(
\zeta_n)\right|_0^1-\frac{\cos^2 \! \!\phi}{\coa^2}
{J}_{(0,\tau+2)}\, .
\end{align}
Splitting,
\begin{align}
 {J}_{(2,\tau)}^{[z]}=&\frac{\beta^2}{\coa^4}
{J}_{(0,\tau)}^{[z]}-\frac{\Lambda_d \beta}{\coa^2}
\hat{K}_{d/2+\tau+1}(z) -\frac{2i \beta \cos \! \phi }{ \coa^3}\,{
J}_{(0,\tau+1)}^{[z]} -\frac{1}{\coa^2} {J}_{(0,\tau+1)}^{[z]}+
\frac{i \Lambda_d \cos \!\phi }{ \coa^3} \hat{K}_{d/2+\tau+2}( z)
-\frac{\cos^2 \!\!\phi}{\coa^2} {J}_{(0,\tau+2)}^{[z]}\nn \\
 {J}_{(2,\tau)}^{[z']}=&\frac{\beta^2}{\coa^4} {J}_{(0,\tau)}^{[z']}+ \Lambda_d \frac{1+\beta}{\coa^2}
e^{i \coa \cos \! \phi} \hat{K}_{d/2+\tau+1}( z') -\frac{2i \beta
\cos \!\phi }{ \coa^3}\,{J}_{(0,\tau+1)}^{[z']} -\frac{1}{\coa^2}
{J}_{(0,\tau+1)}^{[z']}-\nn \\& - \frac{i \Lambda_d \cos \!\phi }{
\coa^3}\, e^{i \coa \cos \! \phi } \hat{K}_{d/2+\tau+2}( z')
-\frac{\cos^2 \!\!\phi}{\coa^2} {J}_{(0,\tau+2)}^{[z']}\, .
\label{reduct2_3}
\end{align}
Substituting (\ref{reduct02}) into (\ref{iii2a}) gives the full
expansion of vectorial integral in terms of Macdonald functions.

\section{Integration over frequencies and angles}
\label{angleints}
(a) Here are computed the angular integrals of the general form
\begin{align}\label{jj0}
V_{m}^n=\int\limits_0^{\pi}\frac{\sin^n
\vartheta}{(1-v\cos\vartheta)^m}d\vartheta
\end{align}
with integers $m, n$, which were encountered in the text. For
$2m>n+1$ one finds in the leading order \cite{GKST-2}
\begin{align}\label{jj2q}
V_{m}^{n}=  \frac{   2^{m-1}   \Gamma
 \left(\frac{n+1}{2}\right)
 \Gamma \left(m-\frac{n+1}{2}\right)}{ \Gamma(m)} \,\gamma^{2m -n-1} \,.
\end{align}
For $2m<n+1$ one obtains
\begin{align}\label{jj3}
V_{m}^{n}=
           \frac{2^{n-m}   \Gamma\left(\frac{n+1}{2}\right)
           \Gamma\left(\frac{n+1}{2}-m\right)} { \Gamma(n -m+1
           )}\, .
\end{align}
In the case
  $2m=n+1$ the behavior of the integral is logarithmic. The only one of those needed here is the:
\begin{align}\label{jj4}
V_{2}^{3}= \frac{2}{v^3}\left[\ln \frac{1+v}{1-v} -2v\right]=
4(\ln 2\gamma -1)+\mathcal{O}(\gamma^{-2})\, .
\end{align}
(b) Integration over the remaining angles is performed
next. Since, with our choice of the coordinate system only
one extra angle $\varphi$ enters  the amplitudes, only the following formula is needed:
\begin{align}\label{intSN}
\int_{S^{D-3}} |\sin \varphi|^N d \Omega = \frac{2 \pi^{ (D-3)/2
}} {\Gamma\left(\frac{D-2+N}{2}\right)}
\Gamma\left(\frac{{N+1}}{2}\right)\, .
\end{align}
\vspace{0.2cm}
(c) Finally, computation of the integrals over the frequency or
over the impact parameter involving two Macdonald functions of the
same argument is performed using the formula  \cite{Proudn}:
\begin{align}
\label{intfreq}
 \int\limits_0^{\infty}K_{\mu}(cz)K_{\nu}(cz)z^{\alpha-1}dz=
\frac{2^{\alpha-3}\Gamma \left(\frac{\alpha+\mu+\nu}{2} \right)
\Gamma \left(\frac{\alpha+\mu-\nu}{2} \right)\Gamma
\left(\frac{\alpha-\mu+\nu}{2} \right)\Gamma
\left(\frac{\alpha-\mu-\nu}{2} \right)}{c^{\alpha} \Gamma(\alpha)}
\, .
\end{align}

\section{Asymptotic behavior of retarded fields}
\label{notes}

In this Appendix we derive the asymptotic behaviors of the retarded solutions of the gravitational
wave equation in arbitrary dimensions and of their space-time derivatives.

The derivation is based on the well known formula
\cite{Vladimirov} for the retarded Green's function of the
$D-$dimensional d'Alembert operator
\begin{align}\label{Greens}
G_D(x) =\left\{%
\begin{array}{ll}
    \ds \frac{1}{2\pi^{D/2-1}} \theta(x^0) \delta^{(D/2-2)}(x^2)\, , & \hbox{$D$=even\, ;} \\
   \ds  \frac{1}{2 \pi^{(D-1)/2}} \theta(x^0)
\left(\frac{d}{dx^2}\right)^{\frac{D-3}{2}}
\!\!\left(\frac{\theta(x^2)}{(x^2)^{1/2}}\right)\,, & \hbox{$D$=odd\, ,} \\
\end{array}%
\right.
\end{align}
with $\delta^{(-1)}(x^2)\equiv \theta(x^2)$ and
$x^2=(x^0)^2-(x^1)^2-...-(x^{D-1})^2$.

First, suppose we have one point particle, moving along the
trajectory $\Gamma$, parametrized by  $z^M(\tau)$ and properly
normilized ($\zt^2=1$). The corresponding source of gravitational
the field on flat background is $T^{MN}(x)=m\int
\zt^M(\tau)\zt^N(\tau)\, \delta(x-z(\tau)) \, d\tau.$ The solution
of the d'Alembert equation $\Box \psi_{\rm ret}^{MN}=- \vk T^{MN}$
is the convolution $\psi^{MN}(x)=G_{\rm ret}\ast T^{MN}$. Let us
denote by $\hat\tau$ the retardation point on the smooth
particle-worldline $\Gamma$ corresponding to a given observation
point $x^M=(t,{\bf x})$. Given $x^M$ and the time-like trajectory
$z^M(\tau)$, $\hat\tau$ is the unique solution (with
$x^0>z^0(\hat\tau)$) of the equation $(x-z(\tau))^2=0$, and
specifies the intersection of $\Gamma$ with the past light-cone of
the observation point $x^M$. Using hats to denote all
corresponding kinematical quantities (e.g. $\hat{\dot{z}}\equiv
\dot{z}(\hat{\tau})$), the \LW  solution becomes \footnote{The
square root derivative, which appears in odd-dimensions, is
defined as the convolution
$$(d/dx)^{1/2} f(x)=-\frac{1}{2\sqrt{\pi}}\int_0^x \frac{dt}{t^{3/2}} f(x-t)\, .$$
}
\begin{align}\label{even}
\psi^{MN}_{\rm ret}(x)=-\frac{\vk m}{4\pi^{D/2-1}}
\left(\frac{1}{2\rho}\frac{\partial}{\partial \hat{\tau}}
\right)^{\frac{D-4}{2}} \frac{\hat{\zt}^{M}\hat{\zt}^{N}}{\rho}\,,
\end{align}
where $$\rho(x)\equiv -\frac{1}{2}\frac{\partial}{\partial
\hat{\tau}}(x-\hat{z})^2= \hat{\zt}\cdot(x -\hat{z})>0\, .$$
Note that in the co-moving at the retardation moment $\hat{\tau}$ Lorentz frame, the quantity
$\rho(x)$ coincides with the spatial distance $|\textbf{x}-\hat{\textbf{z}}|$ between the
observation and the retardation points.

Now, introduce the null vector $c^M=(x^M-\hat{z}^M)/\rho$. It
represents the properly normalized vector from the retardation
point towards the observation point. Introduce in addition the
vector $n\equiv c-\hat{\zt}$, which satisfies $n^2=-1$ and $n
\cdot \hat{\zt}=0$, as a consequence of $\hat{\zt} \cdot c=1$ and
$\hat\zt^2=1$. Thus, $c^M$ is the sum of the orthogonal unit
vectors: $\hat{\zt}^M$, which specifies a time direction, and
$n^M$ which is space-like and determines a spatial direction, i.e.
a point on the sphere $S^{D-2}$, which is the ``basis" of a
null-cone with axis along $\hat{\zt}^{M}$ and apex at $\hat z^M$.
Thus, for an arbitrary point $x^M$ on this light cone, $\rho$ is
the distance of the apex to the center of the sphere which
contains $x^M$, and is also equal to the radius of that sphere.
Thus, going to the wave-zone of radiation emitted by the
accelerated particle corresponds to considering the limit $\rho\to
+\infty$ keeping fixed the quantities $n$, $\hat\zt$, $\hat{\ddot
z}$, etc, which are related only to direction and refer to the
apex $\hat z$ of the light-cone.

Now with the help of the relations \cite{React} {\arraycolsep=0.5cm
\begin{align}\label{diffret}
\begin{array}{lll}
\partial_{M} \hat{ \tau} =c_{M}  & &  \dot{\rho}=(\hat{ \ddot{z}}c) \rho-1
 \\
\partial_{M} \rho =\hat{\zt}_{M}+ \dot{\rho} c_{M} & & \ds \partial_{M} c^{N}=\frac{1}{\rho}\left(\delta _{M}^{N}-
\hat{\zt}_{M}c^{N}-c_{M}\hat{\zt}^{N}-\dot{\rho} c_{M}c^{N}
 \right) \\\multicolumn{3}{l}{
 \partial_{M}\hat{\zt}_{N}=
c_{M}\hat{ \ddot{z}}_{N} \, ,\;\;\;  \text{and similar ones for higher derivatives,}}
  \end{array}
\end{align}}
and consecutive-differentiation rule (\ref{even}), one can check
explicitly that in even dimensions the asymptotic behavior of
$\psi^{MN}$ and its derivatives is
\begin{itemize}
    \item $\ds \psi^{MN}_D=\frac{\psi^{MN}_{\rm Rad}}{\rho^{D/2-1}}+...+
    \frac{\psi^{MN}_{\rm Newt}}{\rho^{D-3}}$, with $\psi^{MN}_{\rm Rad}, \pp, \psi^{MN}_{\rm
    Newt}$ -- tensors depending on $c, \hat{\zt}, \hat{\ddot{z}}, \pp$
    \item  $\ds \psi^{MN, P}_D=\frac{\Psi^{MNP}_{\rm Rad}}{\rho^{D/2-1}}+...+
    \frac{\Psi^{MNP}_{\rm Newt}}{\rho^{D-2}}$ with $\Psi^{MNP}_{\rm Rad}, \pp, \Psi^{MNP}_{\rm Newt}$ --
    tensors depending on $c, \hat{\zt}, \hat{\ddot{z}}, \pp$. The same asymptotic behavior holds for higher order
    space-time derivatives, but not for $\Box\psi^{MN}$, for which one obtains instead
    \item $\ds \Box  \psi^{MN}_D = \frac{\dtilde{ \psi}^{MN}_{\rm Rad}}{\rho^{D/2}}+...+
    \frac{\dtilde{ \psi}^{MN}_{\rm Newt}}{\rho^{D-1}} $\, .
\end{itemize}


In odd dimensions, on the other hand, the solutions of d'Alembert
equations are not known in closed form and one has to rely on
approximate asymptotic expansions. In particular, one can use the
expansion (proven for scalar and vector fields in \cite{Spirin}
and conjectured for the tensor field $\psi^{MN}$)
\begin{align}\label{asymptotic1}
\psi^{MN}_D=\frac{\psi^{MN}_{\rm Rad}}{\rho^{D/2-1}}+
\mathcal{O}(\rho^{-D/2})\, ,
\end{align}
 where $ \psi^{MN}_{\rm Rad}$ again depends on $c^M$ and on the retarded quantities $\hat\zt, \hat{\ddot z}$ etc, but
is expressed as an integral along the particle trajectory from
$\tau'=-\infty$ to $\tau'=\hat\tau$. Taking into account that
$c^{M}$ enters into $ \psi^{MN}_{\rm Rad}$ either in the product
$(\hat{z}-z(\tau'))\cdot c$ or with a free index, one can apply
(\ref{diffret}) to the integral and, using $c^2=0$ to obtain the
same behavior as in even dimensions, namely \footnote{Relations (\ref{Greens}) can be combined into the single formula
$$G_D^{\rm ret}(x_D^2)= \theta(x^0) \left( \frac{1}{\sqrt{\pi}}  \frac{\partial}{\partial y}  \right)^{\frac{D-2}{2}} G_2(y)
 \left. \vphantom{\frac{\partial}{\partial y}}\right|_{y=x_D^2} \, ,$$
 where $G_2(y)= \theta(y)/2$ and $x_D^2=x_0^2-x_1^2-
\pp - x_{D-1}^2$. Already from (\ref{Greens}) one may notice that
the solution $\psi_{D+2}$ of d'Alembert  equation in $D+2$
dimensions is proportional to $\Box\psi_D$. Thus if $\psi_D^{MN}$
scales in the wave-zone as $1/\rho^{D/2-1}$, $\Box \psi_D^{MN}$
should scale as $1/\rho^{(D+2)/2-1}=1/\rho^{D/2}$, in agreement
with (\ref{asymptotic2}). }:
\begin{align}
\label{asymptotic2}
\Box \frac{\psi^{MN}_{\rm
Rad}}{\rho^{D/2-1}}=\mathcal{O}(\rho^{-D/2})\,  \quad {\rm and} \quad
\Box\psi^{MN}_D=\mathcal{O}(\rho^{-D/2})\, .
\end{align}
For full gravity these properties do  not hold exactly due to
non-linearities, but to the order discussed here we have effective
linear theory with linear operator $\Box$ and sources $\un T$,
$\un T'$ and $S(\un h, \un h')$, respectively. The first two are
local currents and satisfy the assumptions made above. $S(\un h)$,
on the other hand, is not a current generated by the point-like
source, but, expressed
 explicitly with the help of (\ref{natag}), and taking into account the field equation
(\ref{psi2eq}), one can easily check that this part of $ \Box \de
\psi_{MN}$ falls-off much faster at large distances, namely as the product of two Coulomb
fields,
\begin{align}\label{asymptoticD}
\Box  \de \psi_{MN}(S) =-\varkappa_D S_{MN} \sim \un h \un h'
=\mathcal{O}(\rho^{-(2D-6)}).
\end{align}
Thus, when substituted to replace $\Box  \de \psi_{MN}$ into the integrals
$$
\int\limits_C \de h^{PQ}\, \Box \de\psi_{MP} \,d\sigma_Q \qquad
\text{and} \qquad \int\limits_C \de h\,\Box \de\psi_{MP}\,
d\sigma^P
$$
encountered in Section 2, the resulting integrals vanish as a result of the asymptotic behaviors
(\ref{asymptotic2}) and (\ref{asymptoticD}).
In other words, $S_{MN}$ is significant only in the space-time volume in which both $\un h$ and
$\un h'$ are important, i.e. close to the moment of collision. This justifies the use of the asymptotic
expansions presented above.

\end{document}